\PassOptionsToPackage{framemethod=TikZ}{mdframed}
\documentclass[preprint,authoryear,3p]{elsarticle}

\usepackage{amsmath}
\usepackage{amsfonts}
\usepackage{mathtools}
\usepackage{caption}
\usepackage{subcaption}
\usepackage{graphicx}
\usepackage{algorithm,algorithmic}
\usepackage{amssymb}
\usepackage{svg}
\usepackage[hidelinks]{hyperref}
\usepackage{booktabs}
\usepackage{xcolor} 
\usepackage{tcolorbox}
\definecolor{kulBlue}{RGB}{0,64,122}
\usepackage{siunitx}
\usepackage{pgf} 
\usepackage[colorinlistoftodos]{todonotes}
\DeclareMathOperator\artanh{artanh}
\usepackage{mdframed}

\floatname{algorithm}{Algorithm}
\newcommand{\KwIn}[1]{\textbf{Input:} #1}
\newcommand{\KwOut}[1]{\textbf{Output:} #1}

\newcommand{\mat}[1]{\ensuremath{{\mathbf{\MakeUppercase{#1}}}}}
\newcommand{\R}{\ensuremath{\mathbb{R}}}
\makeatletter
\renewcommand{\vec}[1]{%
	\ifcat\relax\noexpand#1%
	\ensuremath{\boldsymbol{\lowercase{#1}}}%
	\else
	\ensuremath{\mathbf{\lowercase{#1}}}%
	\fi
}
\makeatother

\newenvironment{highlightbox}[2]{%
		\tcolorbox[colback=white,colframe=kulBlue,fonttitle={\bfseries},title=#1,phantomlabel=#2,left=2pt,top=1pt, bottom=1pt,right=2pt]}%
{\endtcolorbox}

\begin{document}

\begin{frontmatter}

\title{Modeling and Interpreting Correlations, Null Distributions and Significance Levels in Neural Tracking of Natural Stimuli}

\author[aff1,aff2]{Simon Geirnaert\corref{cor1}}
\ead{simon.geirnaert@kuleuven.be}

\author[aff1]{Alexander Bertrand}

\author[aff2]{Tom Francart}

\author[aff2]{Jonas Vanthornhout}

\affiliation[aff1]{organization={KU Leuven, Department of Electrical Engineering (ESAT), Stadius Center for Dynamical Systems, Signal Processing and Data Analytics}}
\affiliation[aff2]{organization={KU Leuven, Department of Neurosciences, ExpORL}}

\cortext[cor1]{Corresponding author \newline All authors are also affiliated with Leuven.AI - KU Leuven institute for AI.}

\begin{abstract}
	Neural tracking - the time-locking of neural responses to continuous stimuli such as speech, music, and video - is widely used to study how the brain processes natural input. Tracking strength is typically quantified as the correlation between the recorded neural response and the stimulus, decoded and/or encoded through data-driven models, and this correlation is routinely used to compare stimulus features, models, or settings. However, its magnitude depends not only on how strongly the brain tracks the stimulus, but also on the statistical properties of the signals being correlated. For example, a smallband speech envelope carrying almost no information about speech content yields among the highest correlations, simply because it is easier to reconstruct. Meaningful interpretation therefore requires comparing each correlation to its null distribution: the correlations expected without any stimulus-response relationship. We show that the randomization procedures commonly used to construct this null distribution are not interchangeable: each implicitly encodes a different null hypothesis, and we motivate stimulus-response misalignment as the most practical and appropriate choice. Because reliable null distributions require many permutations, we introduce a semi-parametric model using the normal distribution after the Fisher transform that yields accurate significance levels from only 3-5 min of data and predicts them across analysis window lengths. Building on this, we propose the null-normalized tracking score (NNTS), an interpretable measure placing features and models on a common scale, which relates directly to the widely used match-mismatch accuracy. Applied to EEG from 121 participants listening to continuous speech, represented across eight acoustic and linguistic features, the framework reverses conclusions drawn from raw correlations, providing an efficient and principled methodology for interpreting neural tracking correlations and comparing features and models.
\end{abstract}


\end{frontmatter}

	\section{Introduction}
	\label{sec:intro}
	When attending to a naturalistic, continuous stimulus, the brain's neural responses time-lock to that stimulus, a phenomenon known as \emph{neural tracking}~\citep{aiken2008human,lalor2010neural,ding2012neural,pasley2012reconstructing}. It is typically observed using neurorecording modalities with high temporal resolution, such as electroencephalography (EEG), electrocorticography (ECoG), or magnetoencephalography (MEG), and has been demonstrated across a range of naturalistic stimuli, including natural speech, music, and natural video. Unlike traditional neuroscientific and brain-computer interface applications based on synthetic, repeated stimuli, neural tracking of naturalistic, continuous stimuli enables a broader range of experiments, analyses, and applications in real-world scenarios. Examples include studying speech processing~\citep{aiken2008human,ding2012neural,pasley2012reconstructing}, music perception~\citep{sturm2015multivariate,diLiberto2020cortical,diLiberto2021accurate,zuk2021envelope}, visual processing~\citep{ki2020visually,yao2023identifying}, attention in cocktail-party scenarios~\citep{osullivan2014attentional,geirnaert2021eegBased,geirnaert2021unsupervised}, and many more.
	
\begin{figure}
		\centering
		\includegraphics[trim={2.2cm 7.5cm 2.2cm 7.25cm},clip,width=\linewidth]{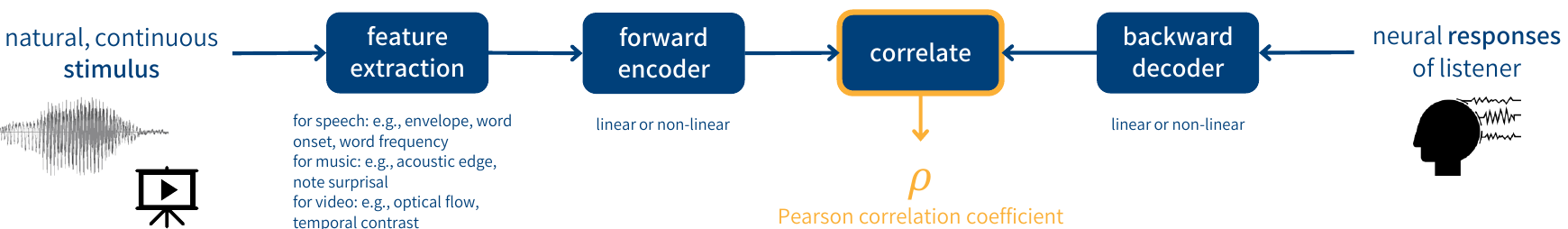}
		\caption{Data-driven models relate natural stimuli such as speech, music, and video to neural responses. Their outputs are typically evaluated using the Pearson correlation coefficient, whose interpretation requires comparison with the corresponding null distribution and significance level.}
		\label{fig:scheme-correlations}
	\end{figure}

	To study and utilize neural tracking across these applications, data-driven models are employed to relate stimulus features to neural responses (Figure~\ref{fig:scheme-correlations}). These models can decode stimuli from neural responses (backward modeling~\citep{wong2018comparison,alickovic2019tutorial,geirnaert2021eegBased}), encode neural responses from stimulus features (forward modeling~\citep{wong2018comparison,alickovic2019tutorial,geirnaert2021eegBased}), or jointly optimize both representations~\citep{dmochowski2018extracting,decheveigne2018decoding,geirnaert2021eegBased,katthi2021deep}, using either linear or non-linear models. Their outputs - or the original signals if no model is used - are typically compared using the Pearson correlation coefficient, the so-called `neural tracking correlation'. This similarity metric is widely used to compare models~\citep{katthi2021deep,thornton2022robust}, stimulus features~\citep{diLiberto2015low,gillis2021neural}, or frequency bands~\citep{diLiberto2015low,etard2019neural,zuk2021envelope,thornton2022robust}, and in several applications serves as a proxy for speech understanding~\citep{vanthornhout2018speech,verschueren2019neural} or auditory attention~\citep{osullivan2014attentional,geirnaert2021eegBased,roebben2024you}. Excellent reviews and tutorials on neural response modeling to natural continuous stimuli can be found in~\citet{crosse2021linear}, with a focus on linear models, and~\citet{puffay2023relating}, with a focus on non-linear models.
	
\begin{figure}
		\centering
		\includegraphics[width=0.9\linewidth]{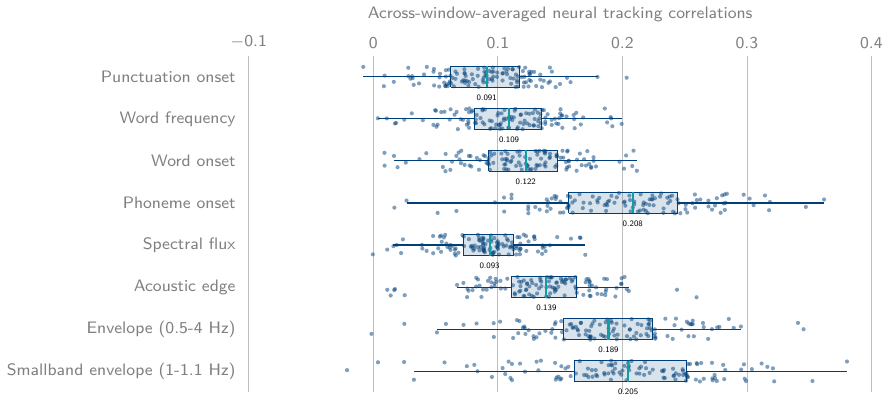}
		\caption{The per-participant average neural tracking correlations (Pearson correlation coefficient) across $\SI{5}{\second}$-windows show clear differences between speech features, with the smallband envelope $(\SIrange{1}{1.1}{\hertz})$ yielding among the highest neural tracking correlations. The high correlation for a feature that carries hardly any information shows that correlation as such is not a proper metric to compare the performance of different features or models.}
		\label{fig:correlations-features}
	\end{figure}

	The main premise of this paper is that, despite its widespread use, the Pearson correlation coefficient is not directly interpretable as a performance metric. Comparing raw correlations across different models, stimulus features, preprocessing strategies, or experimental conditions can therefore lead to wrong conclusions~\citep{zuk2021envelope,puffay2023relating}. As such, terms as `reconstruction accuracy', sometimes used to denote this neural tracking correlation, while intuitive, suggest a well-calibrated performance scale for direct comparison that generally does not exist. Consider, for example, the correlations found in Figure~\ref{fig:correlations-features}, obtained by reconstructing different speech features from EEG using backward modeling. Based solely on the observed correlations, one might conclude that the smallband ($\SIrange{1}{1.1}{\hertz}$) envelope is one of the best features for relating speech to neural responses, influencing subsequent applications in, e.g., attention decoding. However, narrowing the frequency band even further - ultimately approaching a single sinusoidal component - would likely increase the correlation even more, as only the phase needs to be reconstructed accurately. Such a feature, however, carries little information about speech processing or attention. This raises the question of whether higher correlations always indicate better performance and whether the correlation coefficient in itself is a reliable performance metric. This interpretability issue becomes even more pressing in the age of deep learning, where black-box models and their outputs are far less interpretable and controllable. Furthermore, it connects to the finding that canonical correlation analysis (CCA)-based models, which explicitly maximize correlation, tend to favor very low over higher frequency components~\citep{decheveigne2018decoding}. This is expected, as low-frequency signals exhibit stronger temporal autocorrelation, making phase alignment easier without necessarily improving the underlying neural representation. Consequently, maximizing the correlation coefficient alone does not necessarily optimize the quantity of interest\footnote{This also holds for unidirectional backward or forward models, shown to optimize correlation through minimizing the squared error~\citep{biesmans2017auditory}.}.

	\begin{highlightbox}{Key take-away \#1}{}
		The Pearson correlation coefficient between neural responses and stimulus features is \textbf{not} a reliable performance metric \textbf{in itself}. Its interpretation depends on the corresponding null distribution, which can be influenced by various factors as the stimulus feature, model, or preprocessing.
	\end{highlightbox}
	
	To interpret neural tracking correlations meaningfully, they must be evaluated relative to the \emph{null distribution}, representing the scenario in which no relationship exists between the stimulus and neural response. Under the null hypothesis, the expected correlation equals zero, but finite sample sizes inevitably produce non-zero correlations by chance~\citep{combrisson2015exceeding}. Statistical significance is therefore determined from the null distribution, either through a significance level defined as the $100(1-\alpha)\%$-percentile in the null distribution, with typical values for $\alpha = 0.01, 0.05, 0.10$, or even $0.005$ as advocated in \citet{benjamin2018redefine} or through the associated $p$-value, the percentile of the null distribution corresponding to the observed correlation. Although significance is traditionally assessed by comparing the $p$-value with a fixed $\alpha$-threshold, \citet{mcshane2019abandon} advocate instead interpreting $p$-values as continuous measures of evidence, among other factors, rather than binary decision criteria.
	
	Constructing this null distribution is itself a non-trival problem. Historically, standard parametric significance tests, such as the classical t-test, were used as it was impossible to empirically observe the underlying null distribution. However, these parametric tests rely on assumptions such as the independence of the individual time samples entering the correlation computation and a specific parametric distribution - assumptions that are typically violated. Nowadays, however, we have the means and computing power to generate correlations under the null hypothesis based on randomized stimulus-response pairings, for example, using circular shifting or phase scrambling~\citep{crosse2021linear,puffay2023relating}. These non-parametric permutation approaches approximate the null distribution directly, removing the need to assume an underlying parametric distribution. The ideas of permutation testing to determine statistical significance were originally developed in the works of Fisher~\citep{fisher1935design} and Pitman (e.g., \citet{pitman1937significance}). Several works, also in biomedical research, have shown that permutation testing is superior to parametric testing, as it better operationalizes the null hypothesis by translating it directly into a computational procedure~\citep{noguchi2021permutation,holt2023permutation}. However, although permutation tests are commonly described as assumption-free (e.g., \citet{crosse2024permutools}), every permutation procedure implicitly encodes assumptions about the null hypothesis, making the choice of the permutation procedure crucial and non-trivial for obtaining meaningful significance estimates~\citep{holt2023permutation}. Excellent tutorials and overviews of permutation testing can be found in~\citet{maris2007nonparametric,holt2023permutation}.
	
	In this paper, we present a statistically principled framework for using and interpreting the neural tracking correlations, facilitating correct comparison of models, stimulus features, preprocessing, or experimental conditions. This framework addresses three challenges, discussed in the corresponding sections:
	\begin{enumerate}
		\item We analyze commonly used permutation procedures for neural tracking and argue that stimulus-response misalignment provides one of the most principled and practical estimates of the null distribution. Alternative approaches, including random shuffling, circular shifting, and phase scrambling, are shown to rely on different assumptions that can lead to biased significance estimates (Section~\ref{sec:null-dist}).
		\item We introduce a semi-parametric model of the permutation distribution based on the Fisher transform, combining the strengths of non-parametric and parametric modeling, and as such enabling accurate and efficient estimation of significance level and $p$-values. Permutation methods require thousands of permutations to generate reliable null distributions, imposing a heavy computational burden especially when comparing many models, features, or settings. We demonstrate that a parametric model on top of a non-parametric null distribution yields more reliable results than the raw empirical null distribution, especially when only a limited number of permutations is available (Section~\ref{sec:modeling}).		
		\item We combine both contributions into a unified framework for interpreting neural tracking correlations. Building on the modeled null distribution, we introduce the \emph{null-normalized tracking score} (NNTS), a performance metric that enables meaningful comparison across models, stimulus features, and analysis settings, and demonstrate its relationship with the so-called match-mismatch accuracy metric (Section~\ref{sec:use-case}). For this purpose, we revisit the feature-comparison example from Figure~\ref{fig:correlations-features}.
	\end{enumerate}

	Throughout the paper, \emph{``Key take-away''} boxes summarize the main conclusions and can be read independently as a concise overview. A toolbox implementing the proposed methodology is available at~\citet{geirnaert2026correlationToolbox}, and the code and data to reproduce analyses and results at~\citet{geirnaert2026experimentCode} and~\citet{geirnaert2026data}.

	\section{Experimental setup}
	\label{sec:data-experiment-model-details}
	This section describes the experimental setup shared across all analyses in this paper: the stimulus features (Section~\ref{sec:stimulus-features}), the dataset and preprocessing (Section~\ref{sec:dataset}), the backward decoding model (Section~\ref{sec:backward-modeling}), and the validation procedure (Section~\ref{sec:validation-procedure}). As the goal of this paper is to analyze and improve the underlying framework for using and interpreting the neural tracking correlations, and not to benchmark various models or stimulus features, a non-exhaustive list of example features and model is used.

	\subsection{Stimulus features}
	\label{sec:stimulus-features}
\begin{figure} 
		\centering
		\includegraphics[width=0.85\linewidth]{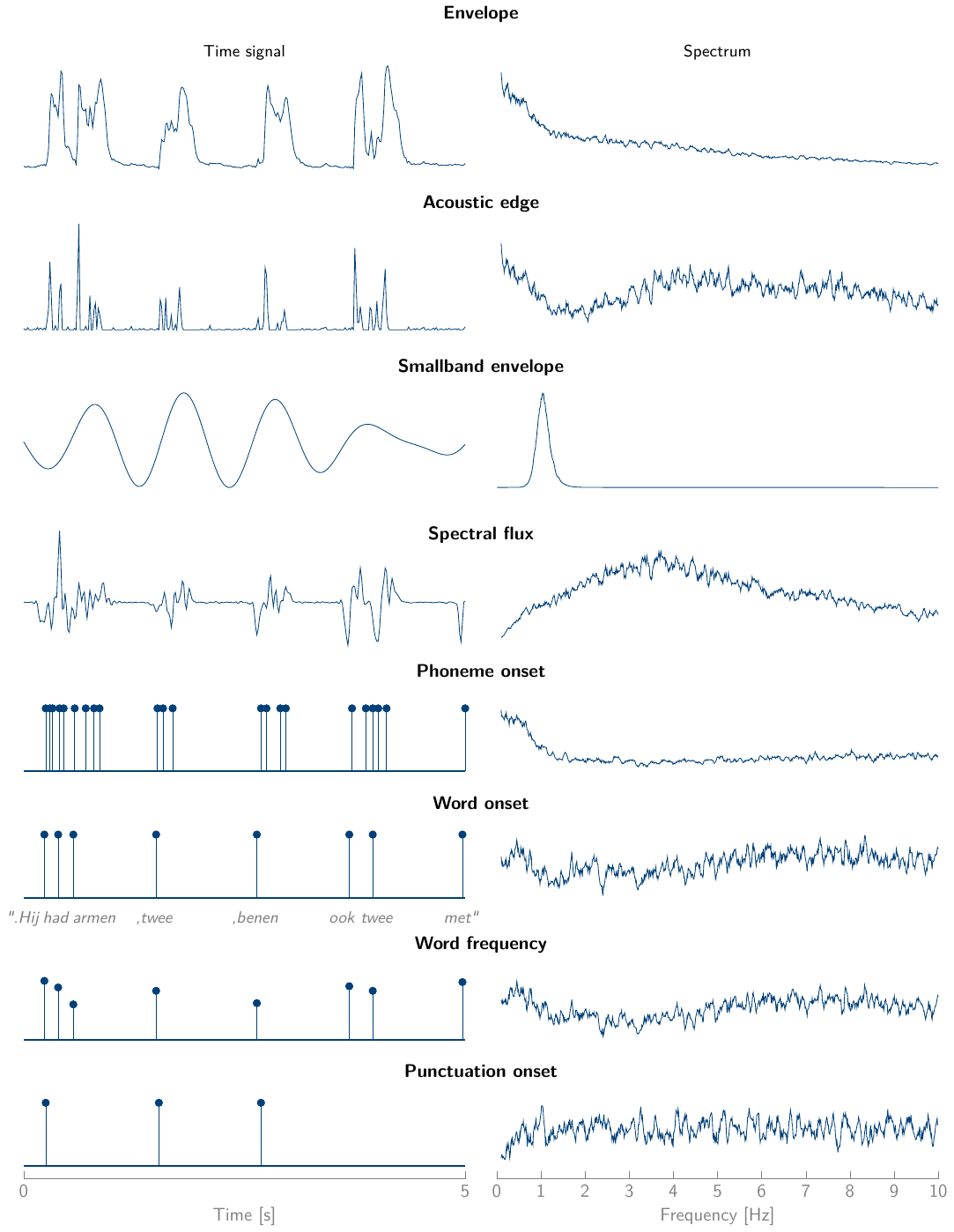}
		\caption{The eight speech stimulus features used throughout this paper, shown for a representative segment in the time and frequency domain (here at $\SI{64}{\hertz}$, shown up to $\SI{10}{\hertz}$ to improve readibility). The continuous acoustic features (envelope, smallband envelope, acoustic edge, spectral flux) are contrasted with the sparse, unfiltered onset-based features (phoneme onset, word onset, word frequency, punctuation onset), which differ markedly in their temporal and spectral statistics.}
		\label{fig:stimulus-features}
	\end{figure}

	We use eight speech features to generate neural tracking correlations, visualized for a representative $\SI{5}{\second}$ segment in Figure~\ref{fig:stimulus-features}:
	\begin{description}
		\item[Envelope] (see, e.g., \citet{ding2012neural}). Generated using a gammatone filter bank with 28 channels spaced by one equivalent rectangular bandwidth, and with centre frequencies from $50$ until $\SI{5000}{Hz}$. Subband envelopes, formed by raising absolute sample values to the power of $0.6$, are averaged into a single envelope. This feature is prevalent in many studies.
		\item[Smallband envelope] The smallband envelope is obtained by filtering the envelope feature between $\SIrange{1}{1.1}{\hertz}$ with a sixth-order Butterworth filter in the forward and backward direction. While this smallband is limited in application, it highlights potential pitfalls in null distribution estimation.
		\item[Acoustic edge] (see, e.g., \citet{brodbeck2018rapid}). Acoustic edges are created by taking the first derivative of the broadband envelope above and setting negative values to zero. They are widely used alongside envelopes in studies focusing on acoustical features.
		\item[Spectral flux] (see, e.g., \citet{macintyre2026decoding}). Spectral flux quantifies the frame-to-frame change in the stimulus' power spectrum, computed as the half-wave rectified sum of the differences between consecutive short-term spectra across the gammatone subbands. Unlike the envelope, it emphasizes spectral changes such as onsets, and is increasingly used as an acoustic feature in neural tracking studies~\citep{macintyre2026decoding}.
		\item[Phoneme onset] (see, e.g., \citet{diLiberto2015low}). Phoneme onsets are created by annotating the stimulus with a value of one at every time step at which a phoneme starts, yielding a sparse impulse train. This feature, or a similar feature carrying more fine-grained information about the phonemes presented, is often used in studies investigating phonetic processing. Importantly, the (near-flat, broadband) spectrum of onset-based features can differ substantially from that of the neural response and of the other features. This spectral mismatch between feature and EEG is innocuous for reconstruction itself, but, as we will see, becomes important when constructing the null distribution.
		\item[Word onset] (see, e.g., \citet{broderick2018electrophysiological}). Word onsets are constructed analogously to phoneme onsets, placing an impulse of value one at the start of each word, and are likewise left unfiltered.
		\item[Word frequency] (see, e.g., \citet{gillis2021neural}). The word frequency feature uses the same word-onset impulses, but modulates each impulse by the log unigram frequency of the corresponding word, so that the feature additionally carries lexical information rather than only timing.
		\item[Punctuation onset] Punctuation onsets, here defined to yield a highly sparse feature, place an impulse at the time of punctuation marks in the narrated text, marking prosodic and syntactic boundaries such as sentence and clause endings.
	\end{description}
	
	\subsection{EEG dataset and preprocessing}
	\label{sec:dataset}
	We used a dataset involving 121 subjects who listened to the story ``Milan'', written and narrated in Flemish by Stijn Vranken. The stimulus is 876 seconds long and was presented monaurally at 60 dBA. Stimuli were presented using the APEX 3 platform on a Windows laptop connected to an RME Multiface II sound card (RME, Haimhausen, Germany) and Etymotic ER-3A insert phones (Etymotic Research, Inc., IL, USA), which were electromagnetically shielded using CFL2 boxes (Perancea Ltd., London, UK). The setup was calibrated in a 2-cm³ coupler (Brüel \& Kjær 4152) using stationary speech-weighted noise matched to the speech material. Experiments took place in an electromagnetically shielded and soundproofed room. The EEG was recorded using a BioSemi 64-channel ActiveTwo system. The dataset used here is part of the publicly available SparrKULee dataset~\citep{accou2024sparrkulee}, yet also contains data from additional participants that could not be made publicly available. All participants, including the additional ones, read and signed an informed consent approved by the Medical Ethics Committee UZ KU Leuven/Research (KU Leuven, Belgium), with reference S57102.

	All EEG data were re-referenced to the common average and drift-filtered (first-order Butterworth, cutoff $\SI{0.5}{\hertz}$, applied forward and backward). All EEG and stimulus features were downsampled to $f_s = \SI{8}{\hertz}$ using an anti-aliasing filter. For the smallband envelope condition, the EEG was additionally filtered to the matching $\SIrange{1}{1.1}{\hertz}$ band.

	\subsection{Backward modeling}
	\label{sec:backward-modeling}	
	A linear backward (decoding) model was used to reconstruct each stimulus feature from the neural responses~\citep{crosse2021linear}. Such a decoder is a spatio-temporal filter that linearly combines the EEG signals of all channels and their time-shifted versions to optimally reconstruct the stimulus feature. Let $r(t,c)$ denote the EEG sample at channel $c$ at time $t$, and let $\mat{R} \in \R^{CL \times T}$ be the matrix collecting $T$ EEG samples across all $C$ channels (here: $C = 64$) expanded with $L$ time-lagged copies with lags up to $\SI{500}{\milli\second}$. The reconstruction $\hat{s}(t)$ of the stimulus feature $s(t)$ is then obtained as
	\[
		\hat{s}(t) = \sum_{c=1}^{C}\sum_{\tau=0}^{L-1} g_c(\tau)\, r(t+\tau,c),
	\]
	with $g_c(\tau)$ the decoder weight for the $\tau$\textsuperscript{th} time lag in channel $c$. 
	
	The decoder is estimated by minimizing the mean-squared error between the reconstruction and the actual feature:
	\[
		g = \arg\min_{g}\; \mathrm{E}\!\left\{\left(\hat{s}(t) - s(t)\right)^2\right\},
	\]
	which yields the following closed-form, ridge-regularized least-squares solution:
	\[
		\vec{g} = \left(\mat{R} \mat{R}^{\top} + \lambda \mat{I}\right)^{-1} \mat{R} \vec{s},
	\]
	where $\vec{g} \in \R^{CL}$ is the vector collecting the decoder weights across all channels and time lags, $\vec{s} \in \R^T$ is the vector containing $T$ stimulus-feature samples, $\mat{I} \in \R^{CL \times CL}$ the identity matrix, and $\lambda$ the ridge parameter counteracting overfitting~\citep{crosse2021linear}. This decoder is then used to reconstruct the stimulus feature, after which the Pearson correlation coefficient between the reconstructed and actual stimulus feature is computed across $N$ samples, corresponding to the window length $w = \frac{N}{f_s}$.
	
	Note that a linear decoder is suboptimal for the sparse onset features (phoneme and word onsets, word frequency, and punctuation onset). A backward decoder reconstructs the feature as a weighted sum of continuous, band-limited EEG channels and their time lags, so its output is an inherently smooth, continuously valued signal. An onset feature, in contrast, is mostly zero with occasional unit impulses - a sparse signal that no linear combination of smooth EEG signals can reproduce; at best the decoder yields a low-pass, smeared estimate with broad bumps around the onset times rather than the impulses themselves, which lowers the attainable Pearson correlation. Recovering the sparse train would require a non-linear (e.g., thresholding) output stage that the linear model lacks. We nonetheless deliberately use the same linear model across all features to keep the comparison controlled and isolate the effect of the null-distribution construction, which is the focus of this work.
	
	\subsection{Validation procedure}
	\label{sec:validation-procedure}
	Using leave-one-window-out cross-validation, the decoder was retrained on all but one window and applied to the held-out window to generate its feature reconstruction; this is repeated for every window of a given length, and for every window length of interest. For each window, the Pearson correlation coefficient between the reconstructed (decoded) stimulus feature from the EEG and the original one is computed. For example, for $\SI{5}{\second}$ windows and the $\SI{876}{\second}$-long story, this results in $\num{175}$ true correlations, yielding a robust average correlation estimate per participant, as visualized in Figure~\ref{fig:correlations-features}. The resulting correlations are available at~\citet{geirnaert2026data}.
	
	\section{Constructing null distributions}
	\label{sec:null-dist}
	Considering the Pearson correlation between a (decoded) neural response and (encoded) stimulus feature, the null distribution represents the correlations expected when no relationship exists between the two signals. In the example used here, the signals consist of a reconstructed stimulus feature, decoded from the EEG using a linear decoder, and the actual feature.

	\begin{highlightbox}{The null hypothesis}{}
		There is no relationship between the neural response and the stimulus feature, i.e., the observed correlation is due to chance.
	\end{highlightbox}

	For the null distribution to be correct, it must reflect the testing context, and in particular the statistical properties of the signals entering the correlation. Stimulus features are typically correlated over time, and this temporal structure (equivalently, their autocorrelation or spectrum) must be preserved when generating null correlations. Furthermore, since the decoder shapes the spectrum of the reconstructed signal and, therefore, also influences the measured correlation, it should also be taken into account when constructing the null distribution. This makes the null distribution decoder-specific, with each model assessed against its own null distribution. 

	Mathematically, the null hypothesis states that the population correlation between neural response and stimulus feature is zero. Consequently, the null distribution, which reflects this hypothesis, must itself have a mean correlation of zero. This is a non-trivial requirement, and one worth checking explicitly once the null distribution has been constructed:
	\begin{highlightbox}{Key take-away \#2}{KTA2}
		The null distribution of the correlation coefficients should have a \textbf{mean of zero}, reflecting the null hypothesis of no relationship between neural response and stimulus feature.
	\end{highlightbox}

	A conceptually different family of approaches randomizes the training process itself, rather than the test-time alignment between a fixed decoder's output and the stimulus feature. Such alternative approaches often correspond, in subtle ways, to different null hypotheses. For example, \citet{macintyre2026decoding} train random decoders on misaligned data but test them on aligned data, where a true relationship exists between neural response and stimulus feature. Besides being considerably more computationally intensive, this procedure relocates the null hypothesis to the level of the model itself, testing whether a trained model extracts meaningful patterns beyond its inductive bias. As this conflates neural tracking with the effectiveness of a specific model's training, we instead test for a relationship directly between the reconstructed signal and the stimulus feature.

	\subsection{Comparing different permutation methods}
	\label{sec:comparison-permutation-methods}
	All non-parametric permutation methods construct the null distribution by repeatedly generating \emph{surrogate signals}: signals stripped of any true temporal correspondence with the neural response, but that ideally retain all other relevant statistical properties of the original feature. Each surrogate is then correlated with the reconstruction of the stimulus feature from the neural response, yielding one sample of the null distribution. Crucially, the properties a permutation method holds fixed when generating surrogates define an implicit \emph{null hypothesis}, and these preserved properties must be matched to the statistic under study~\citep{lancaster2018surrogate}. Different surrogate methods therefore do not approximate a single ``true'' null distribution more or less accurately. They operationalize \emph{different} null hypotheses, and we consider there to be no method-independent best choice. Here, we consider four methods (random shuffling, circular shifting, phase scrambling, and misalignment), examine the null each encodes, and motivate our adoption of misalignment for neural tracking of natural stimuli. The modeling framework of Section~\ref{sec:modeling} and \ref{sec:use-case} is agnostic to this choice: it requires only a set of valid null correlations, so any method can be substituted without affecting the subsequent methodology.

	\subsubsection{Comparing the null distributions}
	\label{sec:comparing-null-distributions}
\begin{figure} 
		\centering
		\includegraphics[width=0.65\linewidth]{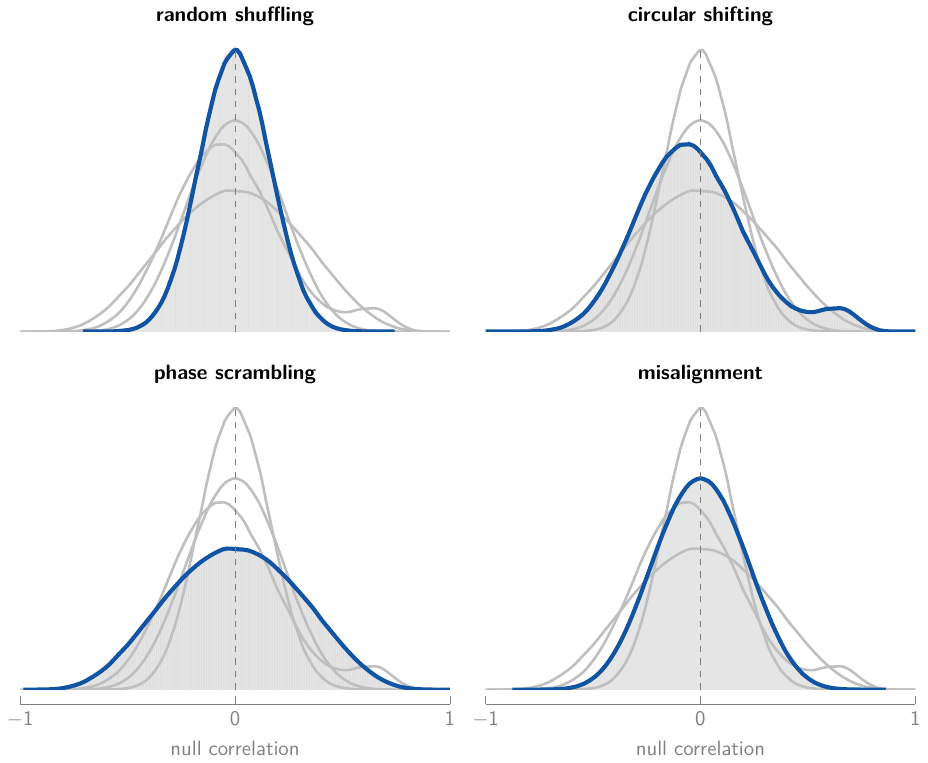}		
		\caption{Null distributions of the 4 permutation methods for the envelope at $\SI{5}{\second}$ windows, across all participants and windows. Each panel highlights one method (histogram and kernel-density estimate) against the other three (gray, identical across panels). Random shuffling whitens the feature, giving a narrow distribution around zero; circular shifting yields a non-zero mean with a secondary bump at high correlations due to autocorrelation; phase scrambling is broad, preserving the amplitude spectrum; and misalignment gives a unimodal distribution reflecting the natural variability of real, temporally unrelated segments.}
		\label{fig:nullDistributionsMethods}
	\end{figure}

	Because each method implicitly operationalizes the null hypothesis in a different way, there is no external ground-truth null distribution against which they can all be validated. Instead, we compare the null distributions the methods produce (Figure~\ref{fig:nullDistributionsMethods}) and reason about the implicit null hypothesis each encodes. The only fully method-agnostic requirement is the necessary condition of a zero mean correlation (Key take-away \#2). The four methods clearly produce different null distributions and therefore different significance levels, confirming that the choice of method is not a mere implementation detail.

	\subsubsection{Random shuffling}
	\label{sec:random-shuffling}
	In random shuffling, the surrogate is generated by randomly permuting the samples of the original (synchronized) feature, i.e., by reordering its values according to a random permutation of the time indices. This implicitly encodes the null hypothesis that the neural response is unrelated to a temporally \emph{unstructured} version of the feature: it destroys any temporal correspondence with the neural response, but also the temporal structure of the feature itself. Permuting the samples whitens the signal and flattens its spectrum: the speech envelope, for example, predominantly contains low-frequency energy, whereas its shuffled surrogate carries substantially more high-frequency energy. Because the correlation between two signals depends on their spectra, and as the whitened surrogate shares little structure with the smooth, low-frequency reconstruction, the resulting null correlations collapse tightly around zero, giving by far the narrowest null distribution (Figure~\ref{fig:nullDistributionsMethods}). This artificially narrow null distribution clearly underestimates what the true spread of a null distribution would be, rendering the test liberal. As its implicit null (no tracking of a whitened feature) corresponds to no meaningful notion of neural tracking, random shuffling is inappropriate for this purpose.

	\subsubsection{Circular shifting}
	\label{sec:circ-shifting}
	In circular shifting, the surrogate is obtained by shifting the original feature over a random number of samples in a circular fashion, such that samples shifted beyond the end of the signal reappear at its beginning. This encodes the null hypothesis that the neural response is unrelated to the feature \emph{at this particular temporal alignment}, testing purely for temporal alignment. It has a clear theoretical appeal: unlike all other methods, it preserves the exact spectrum (and autocorrelation) of the feature. However, precisely because stimulus features can be strongly autocorrelated, small shifts leave the signal - and hence its correlation with the neural response - almost unchanged, producing surrogates that remain partially aligned. In Figure~\ref{fig:nullDistributionsMethods}, this appears as a broad null distribution with a non-zero mean, thus violating the necessary condition of the null distribution, and a secondary bump at high correlation. Conceptually, the null in circular shifting holds the spectrum exactly fixed, testing only alignment rather than whether the response tracks this stimulus content over natural alternatives. Practically (and more decisively), the strong autocorrelation means that, for short windows (typically below a few minutes\footnote{For the envelope, whose autocorrelation extends to approximately $\pm\SI{2.5}{\second}$, yielding \num{100} valid and independent shifts, spaced by the autocorrelation length, requires a window of at least $\SI{4.25}{\minute}$ at $\SI{8}{\hertz}$.}), one cannot draw enough shifts that avoid the autocorrelation to obtain sufficient independent surrogates, making circular shifting practically infeasible.

	\subsubsection{Phase scrambling}
	\label{sec:phase-scrambling}
	In phase scrambling, the surrogate is created in the frequency domain by randomizing the phases (while preserving the conjugate symmetry required for a real-valued signal) and recombining them with the original amplitude spectrum. This encodes the null hypothesis that the feature is a \emph{stationary linear Gaussian process} with a given power spectrum~\citep{lancaster2018surrogate}. This technique removes the practical limitation of circular shifting as it allows generating unlimited independent surrogates from a single short window and avoids the autocorrelation problem. It introduces more randomness than circular shifting, since the phase spectrum changes drastically, but it still exactly preserves the amplitude spectrum, and randomizing the phases destroys the crucial phase-aligned structure (such as acoustic onsets) characteristic of natural auditory features. Various advanced signal processing techniques such as iterative amplitude-adjusted Fourier transform (IAAFT) surrogates have been proposed, that, for example, additionally preserve the amplitude distribution of the feature~\citep{schreiber1996improved}, but still constrain the surrogate to the original power spectrum and therefore share the same fundamental limitation. In Figure~\ref{fig:nullDistributionsMethods}, this yields a broad, smooth null distribution. Like circular shifting, phase scrambling treats the stimulus and/or neural spectrum as fixed and known, which does not reflect the natural variability of stimulus content across a real experiment.

	\subsubsection{Misalignment}
	\label{sec:misalignment}
	In misalignment, the null correlation is computed between the reconstruction $\hat{s}(t)$ and a segment of the \emph{actual} feature taken from a non-corresponding part of the recording (or a different recording with similar characteristics), so that feature and response are misaligned in time. The trained decoder and its reconstruction are kept exactly as in the real analysis, and only the feature against which the reconstruction is correlated is replaced by a genuine but temporally unrelated excerpt. This encodes the null hypothesis appropriate for \emph{content-specific} neural tracking: the neural response is no more related to the presented stimulus than to any other genuine stimulus of the same kind.

	The property distinguishing misalignment from the spectrum-fixing methods is that the feature's statistics such as the spectrum and fine structure are allowed to \emph{vary} across surrogates, because each surrogate is a different real segment with its own statistics. Moreover, by the symmetry of the correlation, the pooled null distribution across windows can equally be interpreted as pairing each stimulus segment with reconstructions from many different epochs, and hence many different background-EEG realizations, such that this variability is present on both the stimulus and the neural-response side. This is important because a valid surrogate should reproduce the variability that would be observed across genuine repetitions of the measurement - a critical property of any surrogate method~\citep{lancaster2018surrogate}. Misalignment reproduces precisely the variability in stimulus-response null pairings that one would expect when the same stimulus is presented repeatedly, rather than treating the stimulus (or neural) spectrum as fixed and known.

	Misalignment does rest on the assumption that the stimulus, and its statistical relationship to the neural response, is approximately stationary across the set of segments from which surrogates are drawn (e.g., the recording). This assumption constrains how the set of misaligned segments should be constructed. Overall, misalignment produces a clean, unimodal null distribution without the pathologies of the other methods (Figure~\ref{fig:nullDistributionsMethods}). Its main practical limitation is that it requires sufficiently long recordings to draw enough misaligned segments, one of the motivations for the semi-parametric modeling introduced in Section~\ref{sec:modeling}.

	Finally, even with misalignment, care must be taken when selecting misaligned segments, to avoid segments that retain non-zero autocorrelation with the original, aligned segment. While circular shifting is vulnerable to autocorrelation at any timescale, even over a few samples, misalignment is vulnerable only to autocorrelation over much longer timescales (seconds or even minutes). One example is fNIRS-based neural tracking~\citep{wilroth2026investigating}, where the very long autocorrelation of the hemodynamic response even causes misalignment to yield spurious correlations, requiring a tailored approach to ensure the null hypothesis is properly satisfied. Long autocorrelation must also be considered when drawing multiple misaligned segments for the same participant, to ensure the resulting null correlations are approximately independent.

	\subsubsection{Summary: misalignment as the adopted default}
	\label{sec:conclusion-misalignment}
	The four methods encode genuinely different null hypotheses and produce correspondingly different null distributions (Figure~\ref{fig:nullDistributionsMethods}). Random shuffling tests against a whitened feature, an uninteresting null that collapses the null distribution and renders the test liberal. Circular shifting and phase scrambling both hold the stimulus spectrum fixed (exactly, or in expectation), testing whether the response tracks this specific realization of the spectrum rather than this stimulus content over natural alternatives. Circular shifting is additionally practically infeasible on short windows below a few minutes due to autocorrelation. Misalignment instead lets the natural statistics of the stimulus vary across surrogates, operationalizing the null hypothesis of content-specific tracking that we here consider most appropriate for neural tracking of natural stimuli. Based on the arguments in Section~\ref{sec:misalignment}, we adopt misalignment throughout the remainder of this paper, with the subsequent framework accommodating any valid alternative.

	\begin{highlightbox}{Key take-away \#3}{KTA3}
		Each surrogate method to create the null distribution encodes a different null hypothesis. We adopt \textbf{misalignment}, as it operationalizes the null hypothesis of \emph{content-specific} neural tracking: it preserves the natural statistics of the stimulus, assuming stationarity, while letting them vary across surrogates, rather than holding the spectrum fixed (circular shifting, phase scrambling) or destroying it (random shuffling). The subsequent framework is agnostic to this choice.
	\end{highlightbox}

	\section{Accurate and efficient semi-parametric modeling of the null distribution and estimation of significance levels}
	\label{sec:modeling}
	In the previous section, we adopted misalignment for generating null correlations. However, reliable null distributions require thousands of permutations, since significance levels are determined by the far tails of the distribution. When data is limited, generating that many misaligned segments may not be feasible. Even with ample data, computing thousands of permutations imposes a substantial computational burden, particularly when null distributions are needed across multiple features, models, or window lengths.
	
	Therefore, we introduce semi-parametric modeling of the null distribution to estimate significance levels more efficiently and accurately. We first introduce the Fisher transformation as the basis for modeling null correlations (Section~\ref{sec:modeling-fisher}), then demonstrate how semi-parametric modeling improves significance level estimation, for example, when permutations are limited (Section~\ref{sec:estimating-si-levels}), and finally show how significance levels can be predicted across window lengths without regenerating the null distribution each time (Section~\ref{sec:predicting-si-levels}).
	
	\subsection{The normal distribution after Fisher transform as a semi-parametric null model}
	\label{sec:modeling-fisher}
	To model the null distribution of Pearson correlation coefficients between neural responses and stimulus features, we use the normal distribution after the Fisher transformation. Fisher proposed a normalizing and variance-stabilizing transformation on the Pearson correlation coefficients $r$ using the inverse hyperbolic tangent function~\citep{fisher1921on}, defined as:
	\[
		z = \frac{1}{2}\ln\!\left(\frac{1+r}{1-r}\right) = \artanh\!\left(r\right).
	\]
	This Fisher transformation maps the correlation coefficients $r \in [-1,1]$ to the real line $z \in ]\!-\!\infty,+\infty[$, such that 
	\[
		z \sim \mathcal{N}\!\left(\mu_z,\sigma_z^2\right)
	\] 
	is approximately normally distributed with mean $\artanh(\rho)$ and variance $\frac{1}{N-3}$, with $\rho$ the true population correlation and $N$ the number of samples in the correlation computation (assuming the $N$ samples are independent). As this variance only depends on $N$ and not on $\rho$, it is called a `variance-stabilizing' transformation \citep{hotelling1953new}. \citet{asuero2006correlation, fouladi2008fisher} showed that this normal approximation holds even for $N$ as low as $11$, making it suitable for modeling null correlations even at short window lengths. In \ref{app:comparison}, we show it outperforms alternatives such as the truncated normal or Student's $t$-distribution.

	For null correlations $r^{(\text{null})}$, where the true population correlation $\rho = 0$ by construction, the mean of $z^{(\text{null})}$ is zero and is modeled as such: $z^{(\text{null})} \sim \mathcal{N}\!\left(0,\sigma_z^{(\text{null})^2}\right)$. Crucially, the variance is not set to $\frac{1}{N-3}$, which assumes independent time samples - an assumption violated by the autocorrelated neural and stimulus signals typical of neural tracking (Section~\ref{sec:null-dist}). Instead, it is estimated empirically from the misalignment permutations $\{z_i^{(\text{null})}\}_{i=1}^{n}$ using the zero-mean unbiased estimator\footnote{Given the mean is assumed zero by construction, no degree of freedom needs to be sacrificed to estimate the mean, resulting in the $\frac{1}{n}$-scaling instead of the generally used $\frac{1}{n-1}$-scaling for unbiased estimation of the variance.}:
	\[
		\sigma_z^{(\text{null})^2} \approx \hat{\sigma}_z^{(\text{null})^2} = \frac{1}{n} \sum_{i=1}^{n} z_i^{(\text{null})^2},
	\]
	with $n$ the number of permutations. This semi-parametric approach retains the key advantage of permutation testing, i.e., correctly capturing the autocorrelation-inflated null variance, while gaining the efficiency of a parametric model for characterizing and extrapolating the tail shape. The significance level at $\alpha$-level can then be derived from this modeled null distribution as
	\begin{equation}
		\label{eq:significance-level}
		\text{significance level}_\alpha = \tanh\!\left(P_{100(1-\alpha)\%}\!\left(\mathcal{N}\!\left(0,\hat{\sigma}_z^{(\text{null})^2}\right)\right)\right),
	\end{equation}
	with $P_x\!\left(\cdot\right)$ the $x$\textsuperscript{th} percentile. The Fisher-transformed normal distribution has been used sporadically in the neural tracking literature, mostly to facilitate $t$-tests on correlation~\citep{fiedler2019late,chen2023speech,lopez2025unsupervised,daeglau2025neural}, but no explicit or systematic semi-parametric modeling of the null distribution using this approach has been proposed.

	To evaluate the semi-parametric model, we compare it against empirical null distributions for the same eight features, using the same data as in Section~\ref{sec:data-experiment-model-details}: the envelope, the smallband envelope, the acoustic edge, spectral flux, phoneme onsets, word onsets, word frequency, and punctuation onsets. Null correlations are generated by misalignment with $\num{100000}$ permutations per participant and window length ($20, 10, 5, 2, \SI{1}{\second}$ at $\SI{8}{\hertz}$)\footnote{For $20$ and $\SI{10}{\second}$ windows, only $\num{9288}$ and $\num{37758}$ permutations, respectively, were used due to the limited number of non-overlapping windows.}, and are treated as the true underlying null distribution. These null correlations~\citep{geirnaert2026data} and the code reproducing all subsequent results~\citep{geirnaert2026experimentCode} are publicly available.

	The semi-parametric model is assessed globally (Section~\ref{sec:modeling-global-assessment}), via fitted PDFs against histograms and QQ-plots, and by how accurately it estimates significance levels from the tail of the distribution (Section~\ref{sec:modeling-si-levels}).

	\subsubsection{Global model assessment}
	\label{sec:modeling-global-assessment}
\begin{figure} 
		\centering
		\begin{subfigure}{0.5\linewidth}
			\centering
			\includegraphics[width=0.8\linewidth]{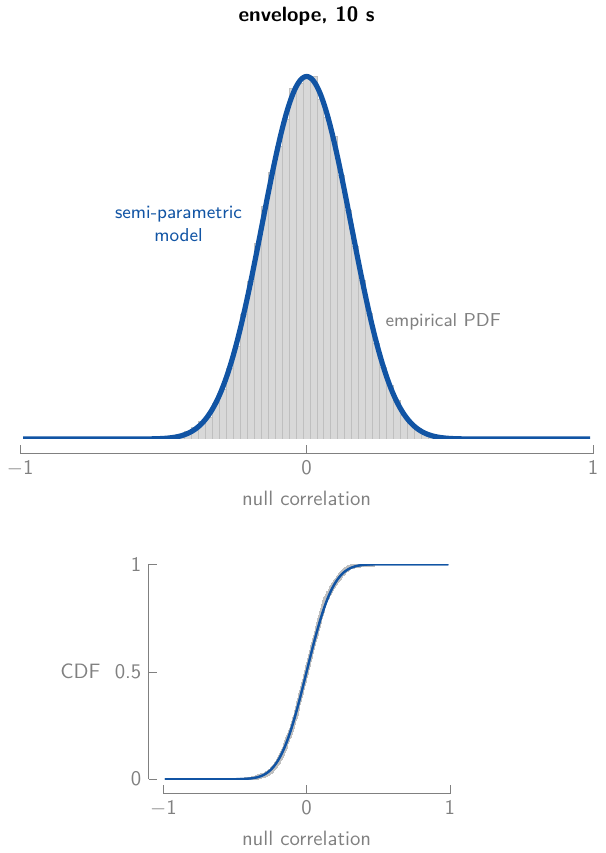}
			\caption{}
			\label{fig:distribution-envelope10s}
		\end{subfigure}%
		\begin{subfigure}{0.5\linewidth}
			\centering
			\includegraphics[width=0.8\linewidth]{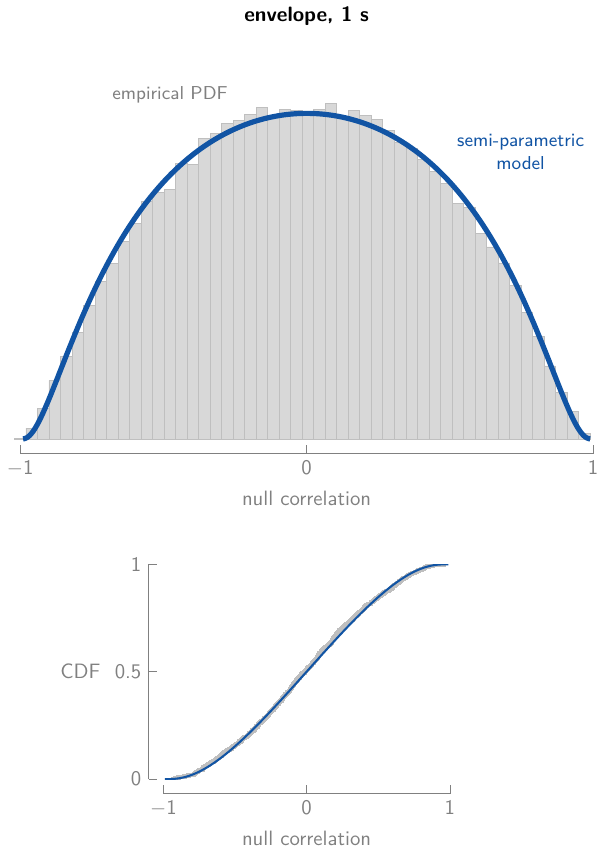}
			\caption{}
			\label{fig:distribution-envelope1s}
		\end{subfigure}
		
		\begin{subfigure}{0.5\linewidth}
			\centering
			\includegraphics[width=0.8\linewidth]{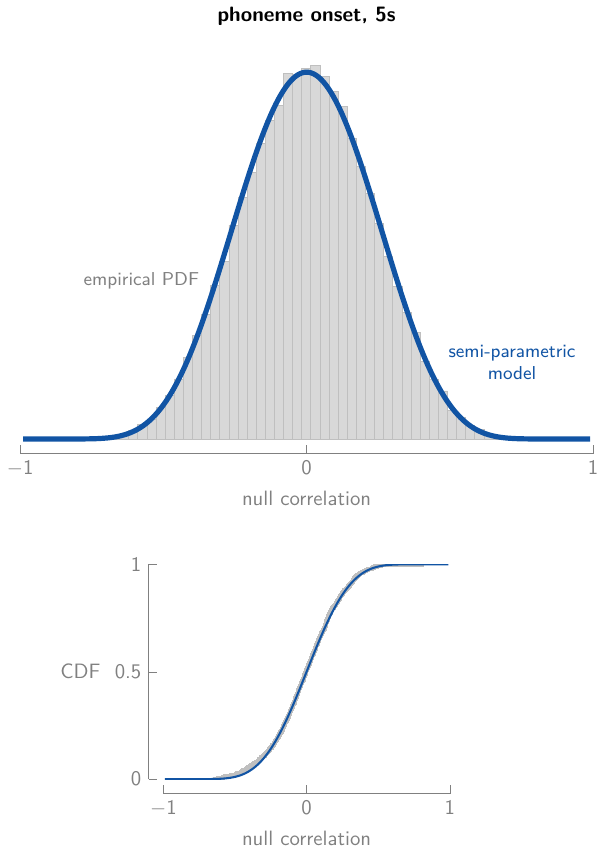}
			\caption{}
			\label{fig:distribution-phonemeOnset5s}
		\end{subfigure}%
		\begin{subfigure}{0.5\linewidth}
			\centering
			\includegraphics[width=0.8\linewidth]{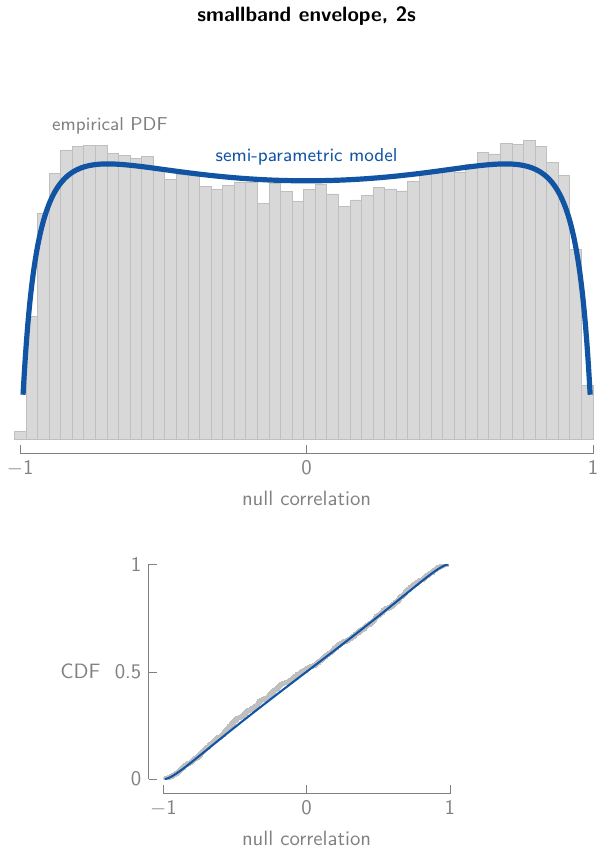}
			\caption{}
			\label{fig:distribution-smallbandEnv2s}
		\end{subfigure}	
		\caption{PDFs and QQ-plots of the semi-parametric model (normal distribution after the Fisher transform) against the empirical null distribution ($\num{100000}$ samples accumulated across all participants) for \textbf{(a),(b)}: the envelope at $\SI{10}{\second}$ and $\SI{1}{\second}$ windows, \textbf{(c)}: phoneme onsets at $\SI{5}{\second}$ windows, and \textbf{(d)}: the smallband envelope at $\SI{2}{\second}$ windows. All panels are shown in the original correlation domain. \textbf{(a),(c)} represent the typical case: the null distribution of raw null correlations closely matches the normal distribution. \textbf{(b)} illustrates the effect of short windows, which widens the distribution and increases non-normality. \textbf{(d)} shows an edge case - short windows combined with a spectrally narrow feature - where heavier tails are present but still well captured by the semi-parametric model.}
		\label{fig:distributionBaseline}
	\end{figure}

	Figure~\ref{fig:distributionBaseline} shows PDFs and QQ-plots of the semi-parametric model against the empirical null distribution (using $\num{100000}$ permutations randomly sampled from the accumulated distribution across all participants) for four representative examples, including some edge cases.

	For typical features and window lengths above $\SI{1}{\second}$ (the large majority of scenarios) the original null distribution already closely follows the normal distribution (Figures~\ref{fig:distribution-envelope10s},~\ref{fig:distribution-phonemeOnset5s}), in which case the influence of the Fisher transform is negligible. This is expected: at longer windows, $N$ is large enough such that correlations remain well within $[-1,1]$ in practice, and the Pearson correlation coefficient, being a normalized sum over many terms and therefore loosely based on the central limit theorem, converges toward normality. In edge cases such as short windows or spectrally narrow features (Figures~\ref{fig:distribution-envelope1s},~\ref{fig:distribution-smallbandEnv2s}), the empirical null distribution has substantially heavier tails, yet the semi-parametric model tracks these faithfully. Since significance levels are determined precisely by the tails, we evaluate this accuracy directly in the next section.
	
	\subsubsection{Significance level assessment}
	\label{sec:modeling-si-levels}
\begin{figure} 
		\centering
		\includegraphics[width=0.6\linewidth]{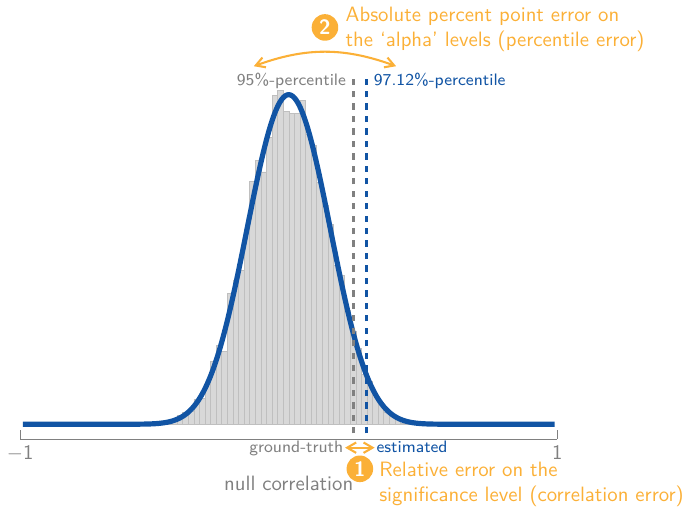}
		\caption{To evaluate the significance level estimation from the modeled distribution, i.e., given a certain $\alpha$-level, what is the corresponding correlation-threshold based on the parametric model, we use two performance metrics. The \emph{correlation error} is the relative error between the estimated and true significance-level thresholds (in correlation units). The \emph{percentile error} is the resulting absolute deviation from the nominal $\alpha$-level in percentage points.}
		\label{fig:performance-metrics}
	\end{figure}

	We evaluate the quality of the semi-parametric model specifically in the context of significance levels, i.e., based on the tail of the null distribution. The estimated significance level from the parametric model for a given $\alpha$-level is assessed using two complementary metrics illustrated in Figure~\ref{fig:performance-metrics}: the \emph{correlation error} (relative error on the estimated significance-level threshold in correlation units) and the \emph{percentile error} (absolute deviation from the nominal $\alpha$-level in percentage points). Both metrics are needed because a small relative error on the correlation threshold does not necessarily translate into a small error on the nominal $\alpha$-level, depending on the probability mass in the underlying distribution.

\begin{figure} 
		\centering
		\begin{subfigure}{1\linewidth}
			\centering
			\includegraphics[width=0.98\linewidth]{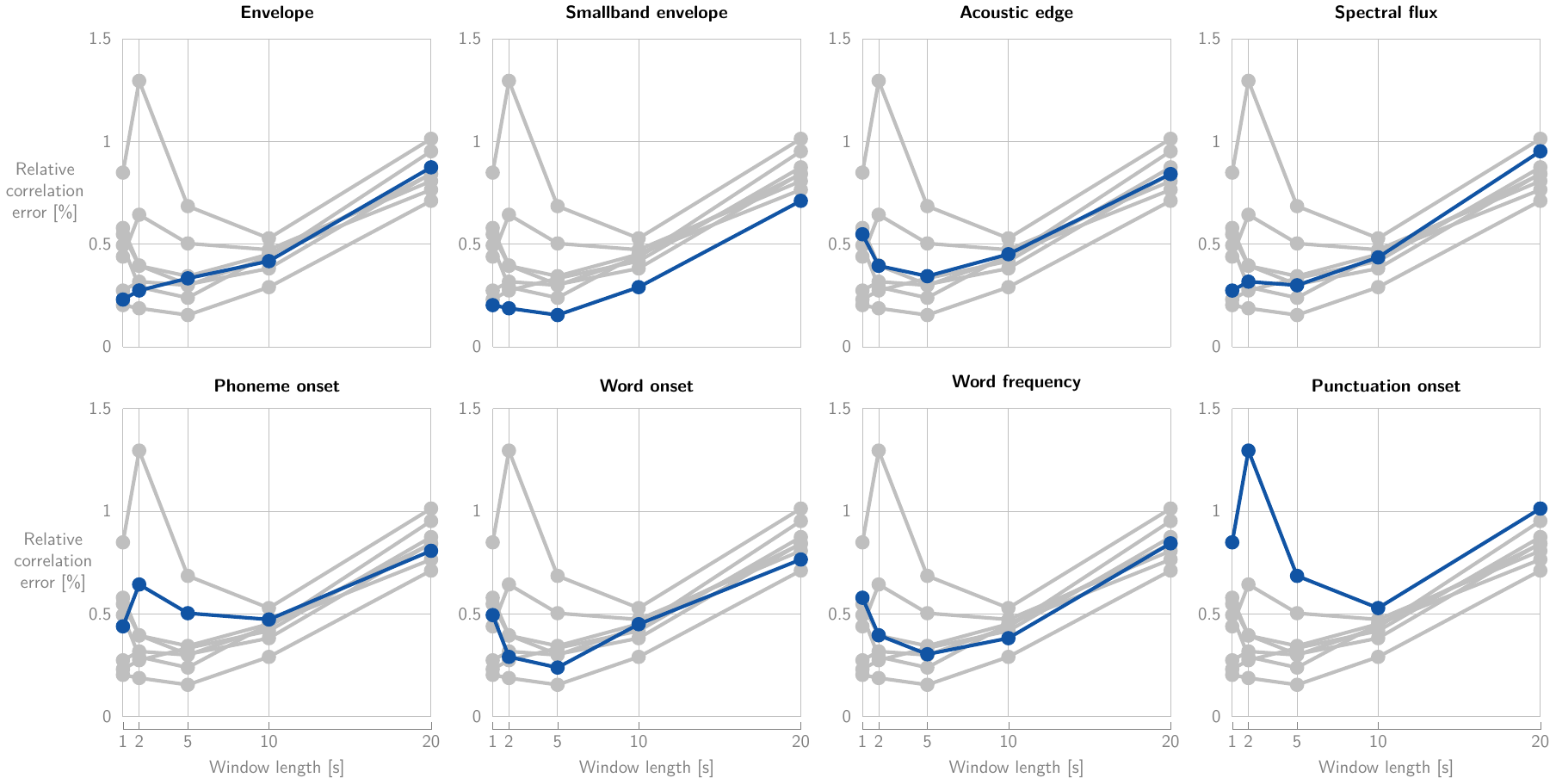}
			\caption{}
			\label{fig:result1_correlationError}
		\end{subfigure}
		
		\begin{subfigure}{1\linewidth}
			\centering
			\includegraphics[width=1\linewidth]{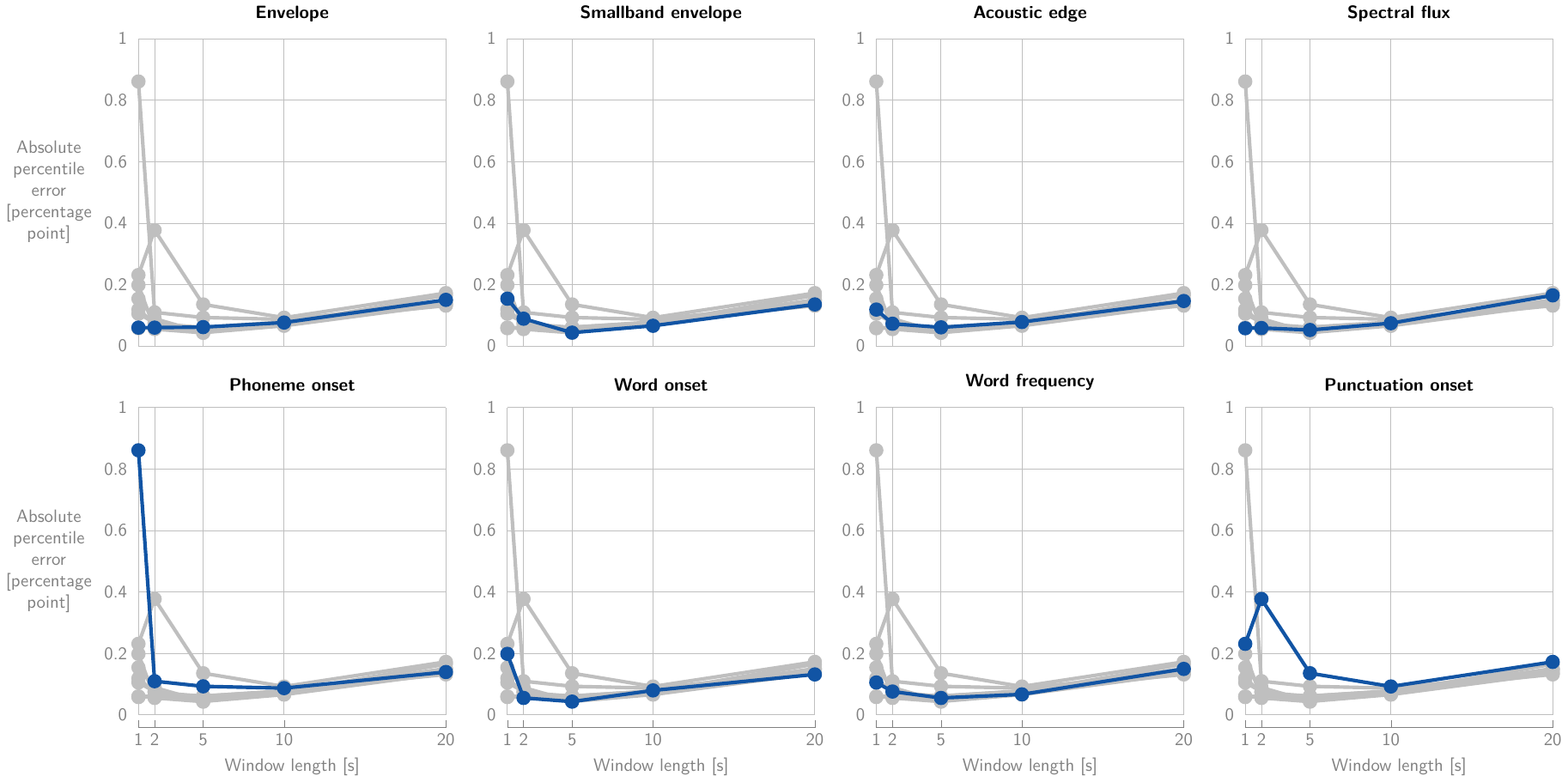}
			\caption{}
			\label{fig:result1_percentileError}
		\end{subfigure}
		\caption{\textbf{(a)} The average (across participants) relative correlation error on the estimated $95\%$-significance level using the normal distribution after Fisher transformation, reminaing below $1.25\%$ across features and window lengths. \textbf{(b)} The average (across participants) absolute percentile error on the nominal $\alpha = 0.05$ level remains below $0.9$ percentage point across features and window lengths.}
		\label{fig:result1_performance-metrics}
	\end{figure}

	Figure~\ref{fig:result1_performance-metrics} shows both metrics per feature and window length, averaged across all $121$ participants. Correlation errors are generally very low (Figure~\ref{fig:result1_correlationError}), averaging $0.51\%$ (standard deviation (SD): $0.26\%$) across all features and window lengths. Percentile errors are equally small (Figure~\ref{fig:result1_percentileError}), averaging $0.13$ percentage points (SD: $0.14$), with the maximum at phoneme onsets and $\SI{1}{\second}$ windows and remaining below $0.5$ percentage points in all other cases. The difference in relative correlation error and absolute percentile error for the phoneme onset feature at $\SI{1}{\second}$ shows the importance of having both metrics: while the error on the significance threshold itself is quite low, in terms of the underlying empirical distribution, the percentile error is larger than for other window lengths due to the wide phoneme onset null distribution. In practical terms, the semi-parametric model estimates the $\alpha = 0.05$ significance level typically with an effective $\alpha$ between $[0.045, 0.055]$.

	These results show that the normal distribution after Fisher transformation yields highly accurate modeling results of the null distribution of correlations and especially of the significance levels found in the tail of these null distributions. Moreover, characterizing the model as a normal distribution with a single estimated parameter, $\hat{\sigma}_z^{(\text{null})}$, brings along several interesting properties and is easy to work with, as shown in the following analyses.
			
	\subsection{Efficient estimation of significance levels}
	\label{sec:estimating-si-levels}
\begin{figure} 
		\centering
		\includegraphics[width=1\linewidth]{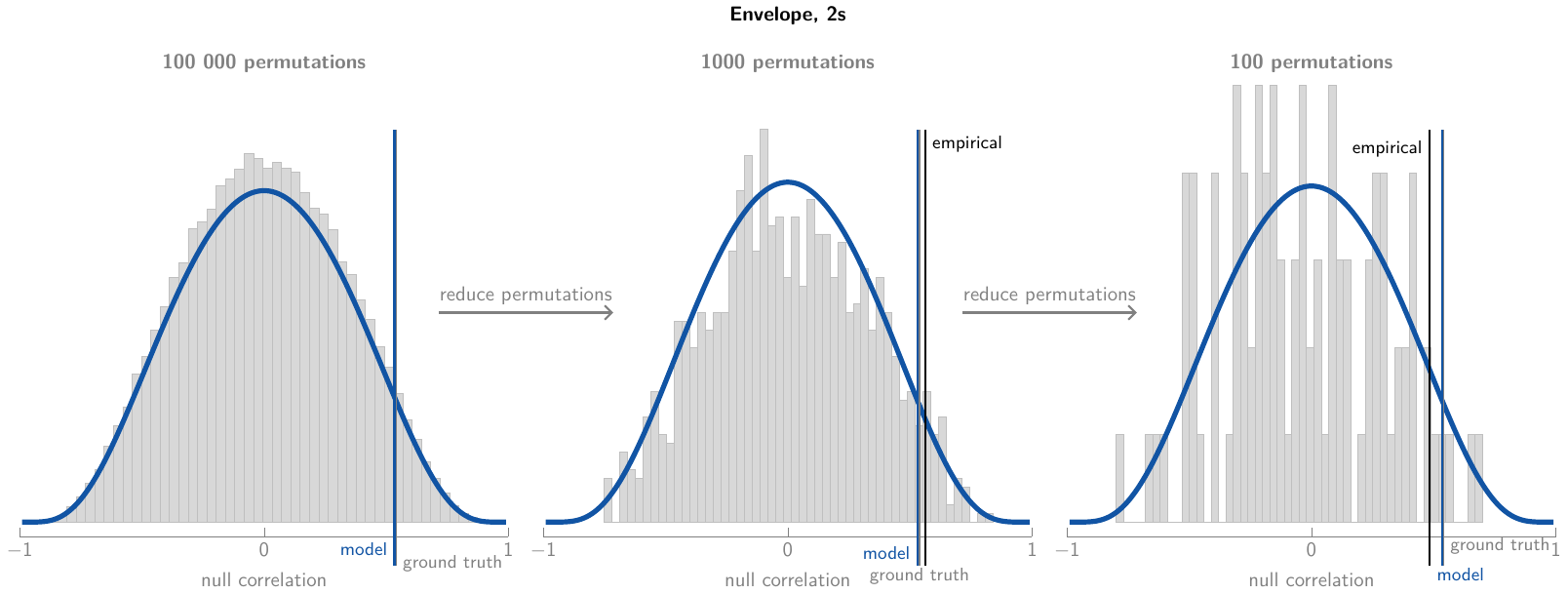}
		\caption{With few permutations, the tail of the null distribution is undersampled, making direct empirical estimation of the significance level, indicated by the vertical line `empirical', unreliable w.r.t. the ground-truth based on \num{100 000} permutations (vertical line `ground truth'). The semi-parametric model mitigates this by modeling the full null distribution based on all available permutations and deriving the significance level, indicated by the vertical line `model', from the analytical tail.}
		\label{fig:example-nbPerms}
	\end{figure}

	Computing $\num{100 000}$ misalignment permutations is computationally expensive, particularly when null distributions are needed across multiple features, models, or window lengths. Depending on the window length, generating sufficient permutations might even be infeasible if data is limited. Yet reliable significance level estimation requires dense sampling of the distribution's tail, which is severely undersampled when few permutations are available, leading to worse significance level estimation based on the empirical null distribution, as illustrated in Figure~\ref{fig:example-nbPerms}.

	The semi-parametric model resolves this tension: rather than estimating the significance level directly from an empirical histogram of null correlations, it models the full null distribution based on all available permutations and extracts the significance level from the modeled, more reliable parametric tail. This pooling of information stabilizes the estimate and reduces its sensitivity to the number of permutations used (Figure~\ref{fig:example-nbPerms}).

\begin{figure}
		\centering
		\begin{subfigure}{0.5\linewidth}
			\centering
			\includegraphics[width=1\linewidth]{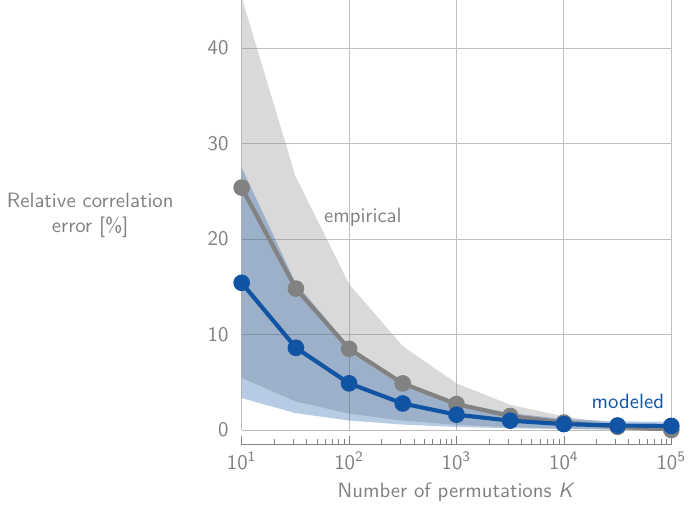}
			\caption{}
			\label{fig:result2_correlationError}
		\end{subfigure}%
		\begin{subfigure}{0.5\linewidth}
			\centering
			\includegraphics[width=1\linewidth]{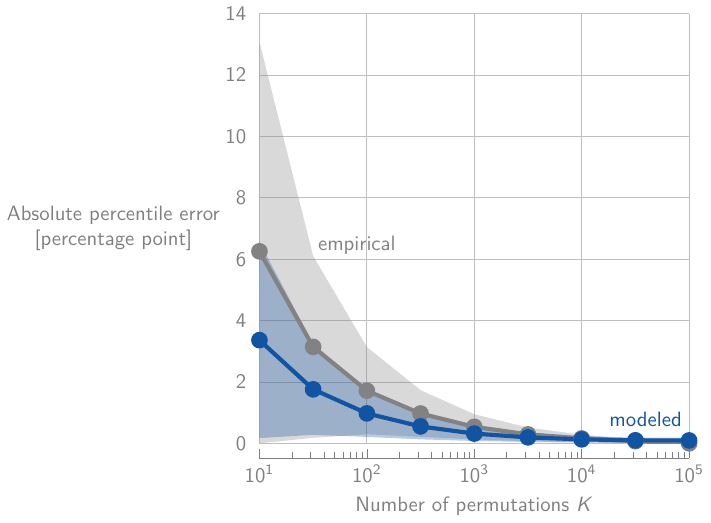}
			\caption{}
			\label{fig:result2_percentileError}
		\end{subfigure}
		\caption{\textbf{(a)} The average (across participants, features, window lengths, and resamplings; shaded area: $\pm 1$ SD) relative correlation error on the estimated $95\%$-significance level as a function of the number of permutations $K$. \textbf{(b)} The corresponding average absolute percentile error shows that the semi-parametric model outperforms the empirical estimate below $K = \num{10 000}$ permutations, with consistently below $1$ percentage point errors when more than $K = \num{100}$ permutations are used.}
		\label{fig:result2_performance-metrics}
	\end{figure}

	To evaluate this, we compare semi-parametric and empirical significance level estimates as the number of permutations is reduced. The null distribution and significance level based on $\num{100 000}$ permutations serves as ground truth. For each combination of number of permutations $K$, feature, participant, and window length, we draw $\num{100}$ random subsets (i.e., $\num{100}$ sets, each with $K$ samples of permuted correlations) and estimate the $95\%$ significance level both empirically and via the semi-parametric model from each $K$-sample subset. Figure~\ref{fig:result2_performance-metrics} shows the resulting average correlation and percentile errors. The semi-parametric model outperforms the empirical estimate for all numbers of permutations below $K = \num{10 000}$ permutations, with a negligible difference at very high numbers of permutations. Notably, the percentile error of the semi-parametric estimate remains below one percentage point on average when at least $K = \num{100}$ permutations are used.

\begin{figure} 
		\centering
		\includegraphics[width=0.7\linewidth]{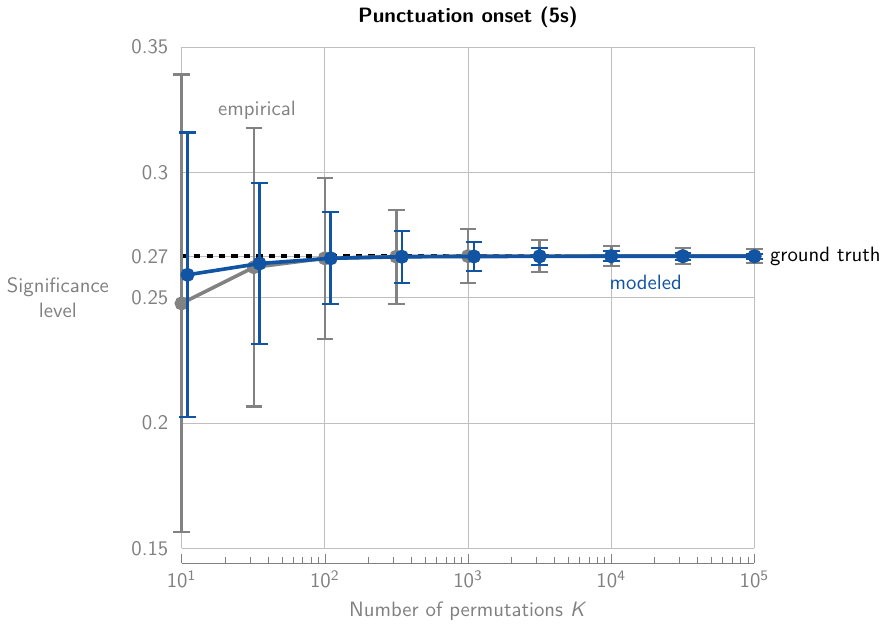}
		\caption{Mean $\pm$ SD (across participants and resamplings) estimated $95\%$-significance level for the punctuation onset feature at $\SI{5}{\second}$ windows, as a function of the number of permutations. The semi-parametric model shows lower bias and substantially lower variance than the empirical estimate, explaining its superior performance.}
		\label{fig:result2_bias-variance}
	\end{figure}

	Figure~\ref{fig:result2_bias-variance} shows the mean $\pm$ SD estimated significance level across participants and resamplings for the punctuation onset feature at $\SI{5}{\second}$ for both the semi-parametric model and empirical estimate across number of permutations $K$, allowing to better explain why the semi-parametric model yields more accurate significance level estimates by analyzing the bias and variance. Firstly, Figure~\ref{fig:result2_bias-variance} clearly reflects that the empirical estimator is a biased yet consistent estimator, as the bias is clearly nonzero for small numbers of permutations, yet asymptotically becomes zero. The semi-parametric model is also biased due to model approximation errors, yet this bias is consistently lower except for very high numbers of permutations where it is negligible. The dominant advantage of the semi-parametric estimate, however, is the variance: across different resamplings, the empirical estimate shows much higher variance than the semi-parametric model, as it must characterize the tail from the few permutations that happen to fall there. In contrast, the semi-parametric model pools all available (limited number of) permutations to fit the full null distribution and extrapolates the tail analytically from a more robust distribution model, yielding much lower variance across a wide range of number of permutations. 

	In summary, the semi-parametric model yields substantially superior significance level estimates up to $\num{10 000}$ permutations, with accurate estimates already from $\num{1000}$ permutations (mean percentile error: $0.32$ percentage points; SD: $0.24$ across participants, features, window lengths, and resamplings).

	\subsection{Extrapolating significance levels across window lengths}
	\label{sec:predicting-si-levels}
	While the previous section shows that the semi-parametric model drastically reduces the number of permutations required for accurate significance level estimation, the entire procedure still needs to be redone for every window length of interest. Here, we propose a method to extrapolate the significance level from one window length to any other, using the normal distribution after the Fisher transform.

	Fouladi and Steiger derived expressions for the first eight cumulants of the Fisher transform under the null hypothesis of zero population correlation~\citep{fouladi2008fisher}. For the variance of $z^{(\text{null})} = \artanh\left(r^{(\text{null})}\right) \sim \mathcal{N}\!\left(0,\sigma_z^{(\text{null})^2}\right)$, this yields the following expression\footnote{Except at very low sample sizes $\leq 10$, this variance is well approximated by $\frac{1}{N-3}$, as proposed in Section~\ref{sec:modeling}.}:
	\begin{equation}
		\label{eq:variance}
		\sigma_z^{(\text{null})^2} = V(N) =  
		\begin{cases}
			\dfrac{\pi^2}{12} - \dfrac{1}{2}\displaystyle\sum_{k=1}^{\frac{N-4}{2}} k^{-2},& N\text{ even} \\
			\\
			\dfrac{\pi^2}{4} - 2\displaystyle\sum_{k=1}^{\frac{N-3}{2}} (2k-1)^{-2},& N\text{ odd},
		\end{cases}
	\end{equation}
	with $N = w f_s$ the number of samples used to compute the correlation, $w$ the window length in seconds, and $f_s$ the sampling frequency in Hz. Crucially, consistent with the explanation in Section~\ref{sec:modeling}, this expression $V(N)$ for the variance \emph{only} depends on $N$, and hence the window length. Following the same approach used for performance curve modeling in selective auditory attention decoding~\citep{geirnaert2025performance}, we use Equation~\eqref{eq:variance} not to compute the variance directly (given the independence assumption is violated), but to \emph{rescale} an already-estimated variance from one window length to another. Given an estimated variance $\hat{\sigma}_{z,\text{base}}^{(\text{null})^2}$ at a base window length $w_\text{base} = N_\text{base}/f_s$, the variance at any target window length $w_\text{target} = N_\text{target}/f_s$ is:
	\begin{equation}
		\label{eq:extrapolated-variance}
		\hat{\sigma}_{z,\text{target}}^{(\text{null})^2} = \hat{\sigma}_{z,\text{base}}^{(\text{null})^2}\frac{V(N_\text{target})}{V(N_\text{base})}.
	\end{equation}
	Combined with Equation~\eqref{eq:significance-level}, the significance level at any target window length $w_\text{target}$ can then be extrapolated from a single estimated variance $\hat{\sigma}_{z,\text{base}}^{(\text{null})^2}$ at $w_\text{base}$:
	\begin{equation}
		\label{eq:extrapolated-SIlevel}
		\text{significance level}_\alpha(w_\text{target}) = \tanh\!\left(P_{100(1-\alpha)\%}\!\left(\mathcal{N}\!\left(0,\hat{\sigma}_{z,\text{base}}^{(\text{null})^2}\frac{V(N_\text{target})}{V(N_\text{base})}\right)\right)\right).
	\end{equation}

\begin{figure} 
		\centering
		\includegraphics[width=0.8\linewidth]{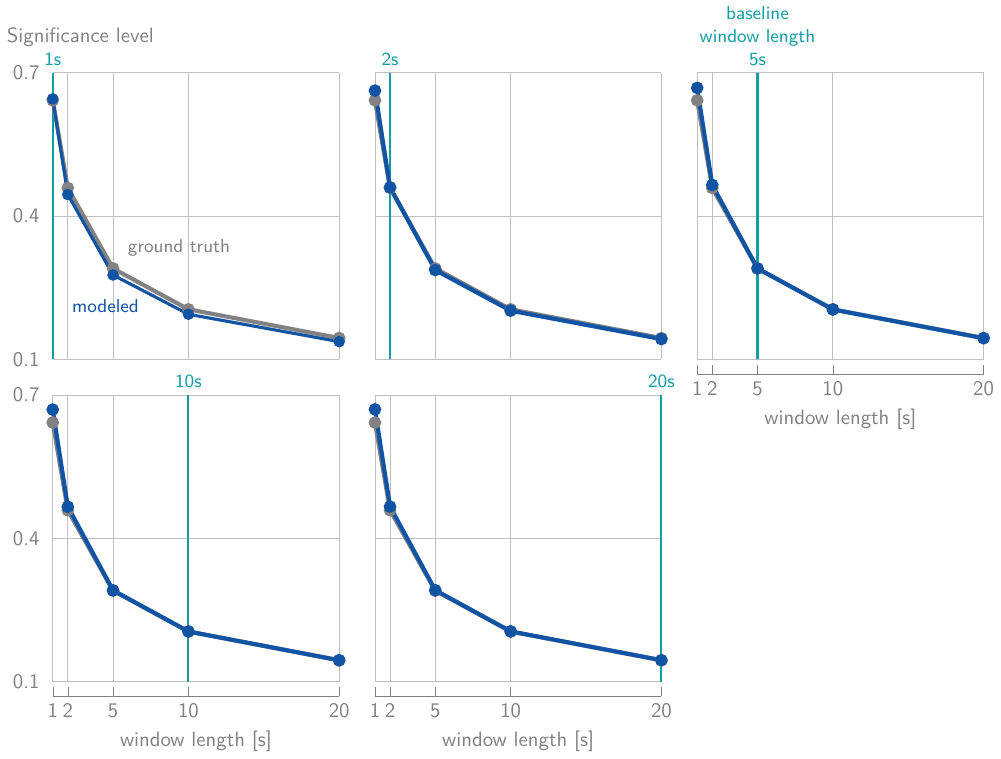}
		\caption{Extrapolating the significance level using the semi-parametric model gives accurate predictions across window lengths, shown here starting from each window length in turn as baseline (shown by the vertical line). The mean significance level (correlation domain) is shown across participants for the word frequency feature.}
		\label{fig:result3_SIlevel}
	\end{figure}

\begin{figure} 
		\centering
		\includegraphics[width=0.85\linewidth]{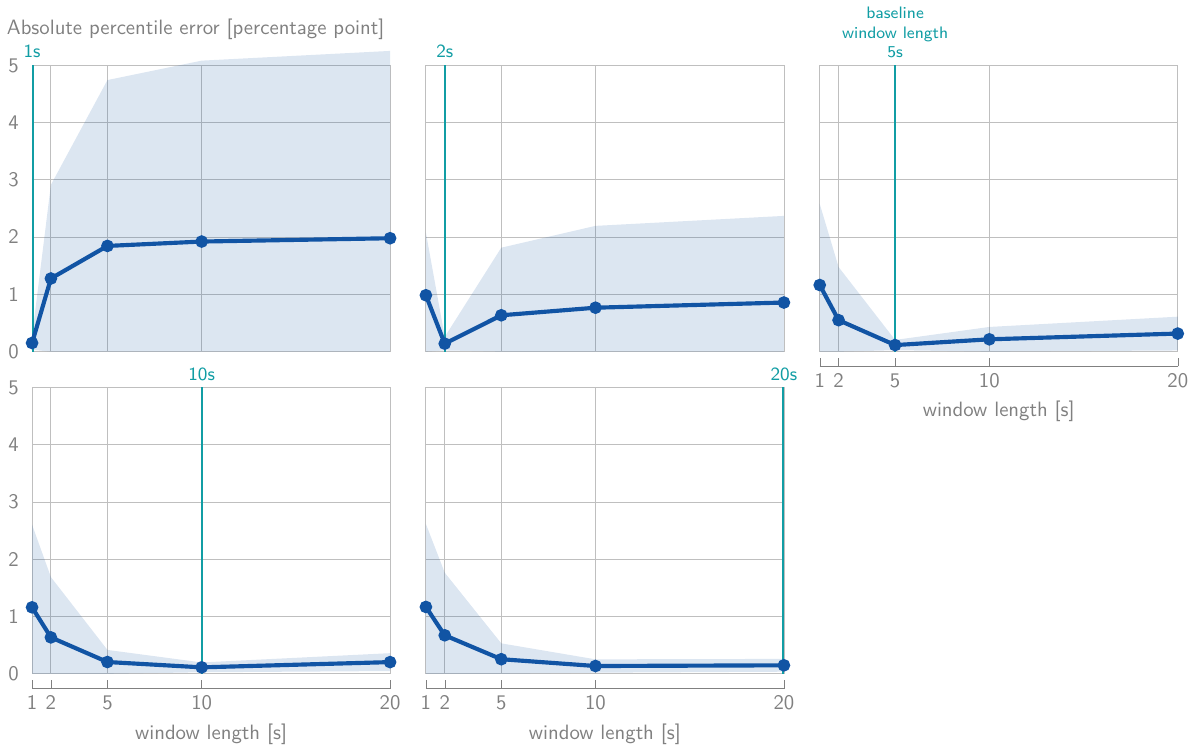}
		\caption{The average (across features and participants; shaded area: $\pm 1$ SD) percentile error, in percentage points, on the extrapolated significance level, starting from various baseline window lengths. As expected, the error is smallest at the baseline window length itself; choosing a baseline near middle of the range, e.g., at $\SI{5}{\second}$, minimizes error overall.}
		\label{fig:result3_percError}
	\end{figure}

	To evaluate extrapolation performance, we extrapolate the significance level between every pair of window lengths, using $\num{10000}$ permutations per participant and feature (sufficient for the semi-parametric model as established in Section~\ref{sec:estimating-si-levels}). Figure~\ref{fig:result3_SIlevel} shows the average significance level across participants for the word frequency feature, taking each window length in turn as the baseline. Overall, the extrapolation is highly accurate. As expected, the error is smallest at the baseline window length and grows with distance from it, most notably when extrapolating to the shortest window length ($\SI{1}{\second}$), where the null distribution deviates most from normality and has the largest variance. A baseline window length near the middle of the considered range therefore performs best; here, $\SI{5}{\second}$. This is confirmed by the percentile error across all features (Figure~\ref{fig:result3_percError}): using $\SI{5}{\second}$ as baseline, the average correlation error remains below approximately $1$ percentage point.

	This extrapolation methodology allows, for example, to predict how long neural responses need to be measured before a given target correlation $\rho_{\text{target}}$, determined a priori for example based on population data, becomes significant. Using Equation~\eqref{eq:extrapolated-SIlevel}, one can sweep $w_\text{target}$ to find the shortest window length for which $\rho_{\text{target}} > \text{significance level}_\alpha(w_\text{target})$, without empirically testing each window length in turn (Section~\ref{sec:use-case-significance-level} gives an example). This is relevant, for example, in clinical applications of neural tracking, where it can serve as an objective measure of hearing or speech understanding. In such contexts, determining in advance (so not online and simply based on population data) the minimum recording length needed to reliably detect tracking is valuable for informing the design of efficient protocols.

	In summary, the semi-parametric model allows the significance level to be predicted at any window length from a null distribution estimated at a single baseline window length (e.g., $\SI{5}{\second}$), with an average percentile error below $1$ percentage point, eliminating the need to populate the null distribution separately at each window length.

	\subsection{Summary: a practical algorithm}
	\label{sec:summary-fisher}
\begin{figure} 
		\centering
		\includegraphics[width=0.6\linewidth]{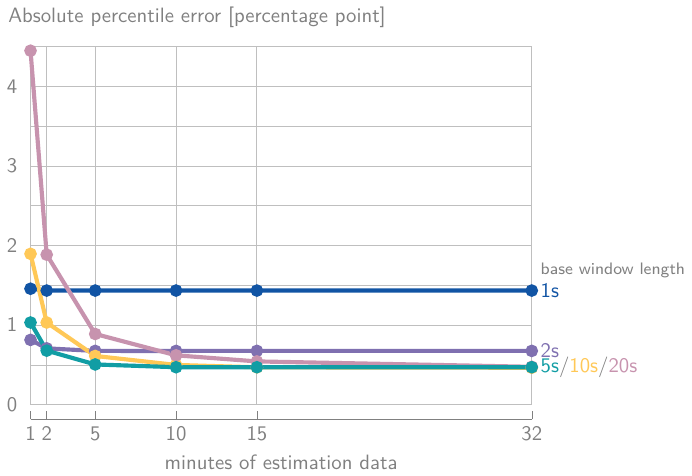}
		\caption{The average (across features, participants, $100$ resamplings, and target window lengths) percentile error, in percentage points, on the extrapolated significance level, as a function of the amount of estimation data available, for various baseline window lengths. For each number of minutes of estimation data, the number of null correlations to estimate the significance level from using the semi-parametric model changes depending on the base window length. Overall, the error is smallest starting from $\SI{5}{\second}$, except when only $\SI{1}{\minute}$ of estimation data is available; using this baseline, the average percentile error remains below $0.5$ percentage points once more than $\SI{5}{\minute}$ of data is available.}
		\label{fig:result3_amountOfData}
	\end{figure}

	In practice, only a limited amount of estimation data is available to generate null correlations via misalignment (Section~\ref{sec:misalignment}), constrained by experiment length and number of trials. This creates a tradeoff: the number of permutations obtainable from a fixed amount of estimation data depends strongly on the baseline window length. For example, $\SI{5}{\minute}$ of estimation data yields more than $\num{10 000}$ permutations at $1$ and $\SI{2}{\second}$ windows, $\num{3540}$ permutations at $\SI{5}{\second}$, but only $\num{870}$ and $\num{210}$ permutations at $10$ and $\SI{20}{\second}$ windows, respectively. With very limited estimation data, it can therefore be preferable to extrapolate the significance level from a shorter baseline window length to the desired target window length, rather than estimating it directly at the target length. Figure~\ref{fig:result3_amountOfData} quantifies this tradeoff, showing the average percentile error on the extrapolated significance level (including at the baseline window length itself) as a function of the amount of estimation data available. Using $\SI{5}{\second}$ as baseline (established above as a good default), the average error stays below $0.5$ percentage points once more than $\SI{5}{\minute}$ of estimation data is available. The resulting procedure using the proposed semi-parametric modeling approach of null distributions, summarized in Algorithm~\ref{algo:siLevel-prediction}, provides an efficient and accurate approach for estimating significance levels across window lengths in neural tracking of natural stimuli. A toolbox, providing among others all code for Algorithm~\ref{algo:siLevel-prediction} is available in~\citep{geirnaert2026correlationToolbox}. 

	\begin{highlightbox}{Key take-away \#4}{}
		Using the normal distribution after the Fisher transform, the null distribution of neural tracking correlations can be modeled such that only $3-\SI{5}{\minute}$ ($\sim \num{1000}$ permutations) of data suffice to reliably and efficiently estimate the significance level across window lengths, features, \dots
	\end{highlightbox}
	
	\begin{highlightbox}{Summarizing algorithm}{}
		\vspace{-\baselineskip}\begin{algorithm}[H]
			\caption{Efficient and accurate estimation of per-window null distributions and significance levels in neural tracking of natural stimuli~\citep{geirnaert2026correlationToolbox}}
			\label{algo:siLevel-prediction}
			\KwIn{(decoded) neural responses and (encoded) stimulus features, sampling frequency $f_s$, target window length $w_{\text{target}}$, $\alpha$-level\\
			\KwOut{null distribution and significance level at the $\alpha$-level for target window length $w_{\text{target}}$}}
			\begin{algorithmic}[1]
				\STATE Use the misalignment method of Section~\ref{sec:misalignment} to generate misaligned pairs of (decoded) neural responses and (encoded) stimulus features. If a single target window length $w_{\text{target}}$ is of interest, set $w_{\text{base}} = w_{\text{target}}$; otherwise, set $w_{\text{base}} = \SI{5}{\second}$. Select the misaligned segments so that they are separated from the aligned segment by more than the autocorrelation length of the signals. Aim for at least $n = 1000$ permutations, yielding empirical null correlations $\{r_i^{(\text{null})}\}_{i = 1}^{n}$. Verify the necessary zero-mean condition: $\bar{r}^{(\text{null})} = \frac{1}{n} \sum_{i=1}^{n}r_i^{(\text{null})} \approx 0$; a substantially non-zero mean indicates residual dependence between misaligned and aligned segments and requires revising the segment selection.
				\STATE Apply the Fisher transformation to the empirical null correlations $\{r_i^{(\text{null})}\}_{i = 1}^{n}
				$: 
				\[
				z_i^{(\text{null})} = \artanh\!\left(r_i^{(\text{null})}\right)
				\]
				\STATE Model the null distribution as a normal distribution by fitting the variance:
				\[
					\hat{\sigma}_z^{(\text{null})^2} = \frac{1}{n} \sum_{i=1}^{n} z_i^{(\text{null})^2},
				\]
				such that
				\[
					z^{(\text{null})} \sim \mathcal{N}\!\left(0,\hat{\sigma}_z^{(\text{null})^2}\right).
				\]
				\STATE Compute the significance level at the $\alpha$-level and at any target window length $w_{\text{target}}
				$ as:
				\[
					\text{significance level}_\alpha(w_{\text{target}}) = \tanh\!\left(P_{100(1-\alpha)\%}\!\left(\mathcal{N}\!\left(0,\hat{\sigma}_z^{(\text{null})^2}\frac{V(N_{\text{target}})}{V(N_\text{base})}\right)\right)\right),
				\]
				with $N_\text{base} = w_{\text{base}}f_s$, $N_\text{target} = w_{\text{target}}f_s$, and
				\begin{equation*}
					V(N) =  
					\begin{cases}
						\dfrac{\pi^2}{12} - \dfrac{1}{2}\displaystyle\sum_{k=1}^{\frac{N-4}{2}} k^{-2},& N \text{even} \\
						\\
						\dfrac{\pi^2}{4} - 2\displaystyle\sum_{k=1}^{\frac{N-3}{2}} (2k-1)^{-2},& N \text{odd}
					\end{cases}
				\end{equation*}
			\end{algorithmic}
		\end{algorithm}
	\end{highlightbox}
	
	\section{Interpreting and comparing neural tracking correlations}
	\label{sec:use-case}
	Building on the null distribution model and significance level provided by Algorithm~\ref{algo:siLevel-prediction}, we return to the use case of comparing different features for neural tracking. We show how the significance level can be used to interpret and compare neural tracking correlations across features (Section~\ref{sec:use-case-significance-level}), and how the proposed null-normalized tracking score (NNTS) and d-prime, together with their connection to match-mismatch accuracy, offer a more principled alternative (Section~\ref{sec:use-case-d-prime}).

	In principle, the null distribution and significance levels could be constructed at three different levels of granularity, always separately per feature (since inter-feature differences are expected to dominate):
	\begin{enumerate}
		\item \emph{Across participants}: all windows across all participants are used to build a \emph{single} null distribution. While this has the advantage of simplicity, resulting in one significance level across all participants that is easy to use, it ignores potential differences in EEG distributions and SNRs between participants.
		\item \emph{Per participant}: a null distribution across all windows is built per participant. This results in one significance level per participant across all windows, which is a bit more complex to work with, yet respects the differences in EEG distributions and SNRs noted above.
		\item \emph{Per individual window}: a null distribution is built for every participant and individual window, resulting in a different significance level per individual window. While this allows taking differences in stimulus properties and non-stationarities in the EEG within participants into account, it is highly cumbersome to work with, and typically insufficient permutations can be generated for reliable per-window null distributions. See also the discussion in Section~\ref{sec:circ-shifting}: to fully exploit per-window characteristics, circular shifting would be the natural choice, yet infeasible in practice.
	\end{enumerate}
	These options trade off convenience against sensitivity to known sources of variability (across features, participants, and individual windows). Here, we adopt the option 2, characterizing a single null distribution - and thus a single significance level - per participant, across all windows. This way, sufficient permutations can be generated per participant, and this granularity naturally aligns with the misalignment technique. Adopting a single per-participant significance level still leaves two ways of using it, both illustrated in the next section: (i) \emph{at the window level}, by comparing each individual window's correlation against this \emph{same}, shared significance level - this is precisely what distinguishes the second option from the third, where every window would instead have its own, window-specific level; and (ii) \emph{at the participant level}, by deriving the corresponding significance level for the across-window-averaged correlation (Algorithm~\ref{algo:avg-estimation}) and testing the participant as a whole.

	The window-level significance level used in~(i) is obtained directly from Algorithm~\ref{algo:siLevel-prediction}. For testing a participant as a whole as used in~(ii), we average the per-window correlations as in Figure~\ref{fig:correlations-features}. Since averaging narrows the null distribution, this requires its own significance level. As shown in \ref{app:model-avg-correlations}, the null distribution of the average (non-Fisher-transformed) correlation across $m$ windows per participant is well modeled by a normal distribution, following directly from the central limit theorem\footnote{Unlike single correlation coefficients, for which the Fisher transform provides an exact normalizing transformation, no equivalent closed-form characterization exists for \emph{averaged} correlation coefficients; here we rely on the central limit theorem instead.}. Its parameters follow directly from the same per-window null correlations $\{r_i^{(\text{null})}\}_{i = 1}^{n}$ used in Algorithm~\ref{algo:siLevel-prediction} (assuming independence between windows, which is typically fulfilled for neural tracking correlations based on $\leq\SI{1}{\second}$ windows~\citep{heintz2025postprocessingeegbasedauditoryattention}), as summarized in Algorithm~\ref{algo:avg-estimation}. An implementation of this algorithm can be found in the accompanying toolbox~\citep{geirnaert2026correlationToolbox}.

	\begin{algorithm}
			\caption{Efficient and accurate estimation of per-participant across-window-averaged null distributions and significance levels~\citep{geirnaert2026correlationToolbox}}
			\label{algo:avg-estimation}
			\KwIn{per-window empirical null correlations $\{r_i^{(\text{null})}\}_{i = 1}^{n}$ on window length $w$, number of windows averaged per participant $m$, $\alpha$-level\\
			\KwOut{across-window-averaged null distribution and significance level at the $\alpha$-level}}
			\begin{algorithmic}[1]
				\STATE Estimate the variance of the per-window raw correlations with mean assumed $0$:
				\[
					\hat{\sigma}_r^{(\text{null})^2} = \frac{1}{n} \sum_{i=1}^{n} r_i^{(\text{null})^2},
				\]
				\STATE Compute the significance level of the across-window-averaged null distribution across $m$ (approximately independent) windows ($\bar{r}_m^{(\text{null})} = \frac{1}{m} \sum_{i=1}^mr_i^{(\text{null})}$, modeled as a normal distribution) at the $\alpha$-level as:
				\[
					\overline{\text{significance level}}_\alpha = P_{100(1-\alpha)\%}\!\left(\mathcal{N}\!\left(0,\frac{\hat{\sigma}_r^{(\text{null})^2}}{m}\right)\right).
				\]
			\end{algorithmic}
	\end{algorithm}

	\subsection{Comparing features via the significance level}
	\label{sec:use-case-significance-level}
\begin{figure} 
		\centering
		\begin{subfigure}{0.9\linewidth}
			\centering
			\includegraphics[width=1\linewidth]{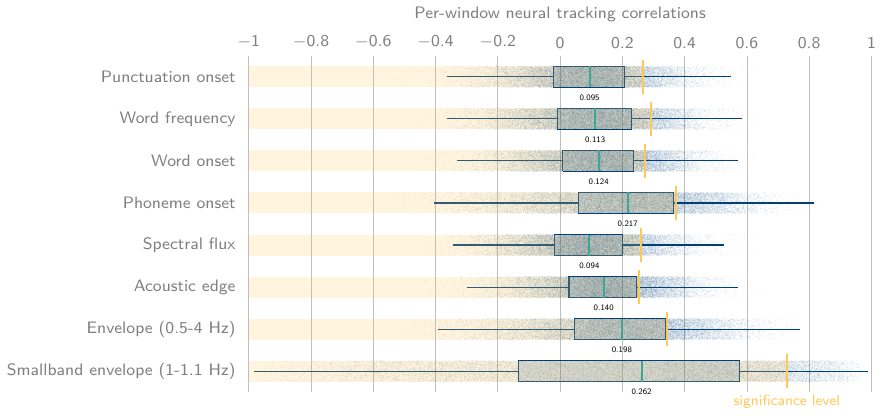}
			\caption{}
			\label{fig:useCase-raw-correlations-withSI}
		\end{subfigure}

		\begin{subfigure}{0.9\linewidth}
			\centering
			\includegraphics[width=1\linewidth]{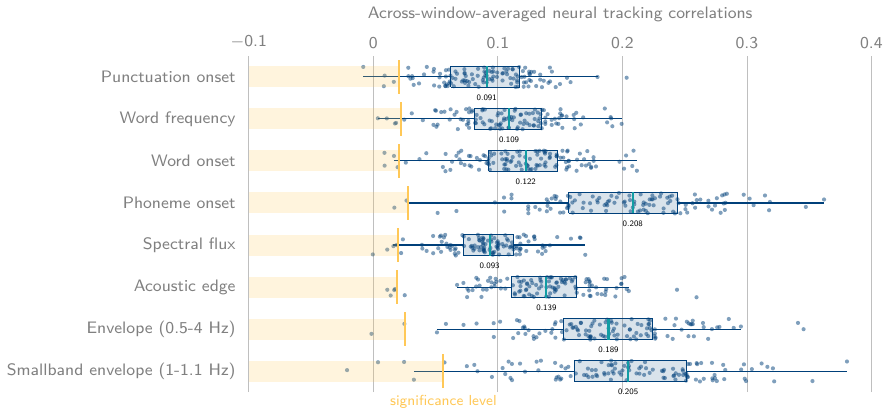}
			\caption{}
			\label{fig:useCase-raw-averaged-correlation-withSI}
		\end{subfigure}
		\caption{\textbf{(a)} The individual neural tracking correlations (one dot = one window) and boxplots across all $\SI{5}{\second}$-windows and participants, for every feature, including the significance levels characterized using Algorithm~\ref{algo:siLevel-prediction}. All correlations below the significance level are non-significance. The per-window significance levels clearly differ across features and show that, for example, smallband envelopes yield, as expected, far fewer significant correlations than other features. \textbf{(b)} The per-participant across-window-averaged correlations (one dot = one participant) and boxplots show how the significance level, characterized using Algorithm~\ref{algo:avg-estimation}, changes when averaging correlations.}
		\label{fig:useCase_raw-correlations}
	\end{figure}

	Figure~\ref{fig:correlations-features} shows, for every feature, the average neural tracking correlations per participant. Without a proper characterization of the null distribution and significance level, one could conclude that the smallband envelope yields the highest neural tracking when interpreting the raw correlation as a direct performance metric. However, computing and plotting the significance level of individual correlations with Algorithm~\ref{algo:siLevel-prediction}, as shown in Figure~\ref{fig:useCase-raw-correlations-withSI}, provides the necessary context for interpreting and comparing neural tracking correlations across features.

	As opposed to Figure~\ref{fig:correlations-features}, Figure~\ref{fig:useCase-raw-correlations-withSI} shows the individual per-$\SI{5}{\second}$-window correlations across all participants, together with the median single-window significance level (usage~(i)). This fully reveals the spread of correlations at a given window length, and is consistent with using Algorithm~\ref{algo:siLevel-prediction}, which characterizes the null distribution at the level of \emph{individual} windows. Testing the across-window-averaged correlations of Figure~\ref{fig:correlations-features} against this same per-window level would be far too conservative, because averaged correlations have a much narrower spread. Applying instead the across-window-averaged significance level (usage~(ii); Algorithm~\ref{algo:avg-estimation}) yields the participant-level test on $\SI{5}{\second}$ windows (averaging $\num{175}$ windows per participant) in Figure~\ref{fig:useCase-raw-averaged-correlation-withSI}: because the null distribution narrows substantially under averaging, the significance level drops sharply. While most individual $\SI{5}{\second}$-window correlations are non-significant (Figure~\ref{fig:useCase-raw-correlations-withSI}), neural tracking can be established at the participant level for all features in most participants (Figure~\ref{fig:useCase-raw-averaged-correlation-withSI}).

	Figure~\ref{fig:useCase-raw-correlations-withSI} shows why the correlation scale is not a good scale for comparing across different features. Because the per-window significance level differs between features, the neural tracking correlations are also \emph{incomparable}. While the smallband envelope might yield the highest correlations, its significance level is also much higher than that of other features, resulting in far fewer windows showing above-significance neural tracking. As explained in Section~\ref{sec:intro}, the smallband envelope feature mostly requires phase alignment on a low-frequency signal, which is much easier to achieve, as now confirmed by its much higher significance level. When the significance level is not the same, the raw correlation is therefore not directly comparable across features. Note that using a forward temporal response function instead of a backward model, predicting neural responses from different features, would place every feature's correlation in the same (EEG) space, which by itself would make significance levels more comparable across features. This does not, however, fully resolve the underlying interpretability problem raised in Section~\ref{sec:intro}: because the significance level still depends on each predictor feature's own autocorrelation structure, features dominated by low-frequency content would still tend toward inflated correlations and inflated significance levels, regardless of model direction.

	As explained in Section~\ref{sec:predicting-si-levels}, using Algorithm~\ref{algo:siLevel-prediction}, we can predict how long neural responses to natural stimuli must be minimally measured before significant neural tracking can be established, or, in other words, what minimum window length should be chosen to establish significance for a given feature. Based on the mean correlation for every feature across windows and participants, we extrapolate the overall $\SI{5}{\second}$-window significance level (based on the global null distribution, with $\num{10000}$ samples accumulated across windows from all participants) to increasing target window lengths, until the mean correlation exceeds the significance level. Implicitly, we assume that the mean correlation remains fairly constant across window lengths, as shown in \citet{lopez2025unsupervised}. The resulting predicted measurement times in Table~\ref{tab:predicted_measurementTimes} show that, for the average participant, the envelope, acoustic edge, and phoneme onset require much less time to establish significant neural tracking (around $\SI{17}{\second}$) than, for example, the smallband envelope (around $\SI{93}{\second}$). Note that these measurement times are only illustrative of the methodology and specific to this dataset and recording setup, rather than general recommendations. Depending on the requirements, more stringent parameters, e.g., significance with a very strict significance level and for most participants in the population, can be inserted.

	\begin{table}[htbp]
		\centering
		\begin{tabular}{r c}
			\toprule
			\textbf{Feature} & \shortstack{\textbf{Predicted measurement time}\\\textbf{to significance [s]}} \\
			\midrule
			Envelope  & 16.8 \\
			Smallband envelope   & 93.4 \\
			Acoustic edge          & 17.4 \\
			Spectral flux           & 38.8 \\
			Phoneme onset & 17.3 \\
			Word onset               & 25.1 \\
			Word frequency            & 35.6 \\
			Punctuation onset          & 43.0 \\
			\bottomrule
		\end{tabular}	
		\caption{Predicted measurement time for the average participant required to measure significant neural tracking for different features, when extrapolating the significance level based on $\SI{5}{\second}$ windows until it is below the mean correlation across participants and windows.}
		\label{tab:predicted_measurementTimes}
	\end{table}

	Based on the correlations (per-window or averaged across windows) and their corresponding modeled significance levels, statistical testing could now be performed: features could be compared, for example, based on the number of correlations above the significance level. However, we advocate against statistical testing in the original correlation space using a fixed significance-level threshold. While the significance level remains a valuable tool for interpreting raw correlations, thresholding at a fixed level inherits all the disadvantages of classical statistical testing. When comparing many windows or participants against the significance level, the question of multiple-comparisons correction arises. For example, when simply using the number of windows or participants yielding above-significance correlations as a quantitative metric, there is a case for omitting corrections; however, once windows or participants are selected or reported based on significance, corrections become necessary. Second, thresholding requires an arbitrarily chosen significance level, discarding all continuous evidence. Better alternatives are therefore to use $p$-values as continuous variables, or NNTS and d-prime (Section~\ref{sec:use-case-d-prime}).

	As a first alternative, the percentile of an observed neural tracking correlation within the null distribution, i.e., its $p$-value, can be used as a continuous measure of neural tracking without thresholding~\citep{mcshane2019abandon}, thereby also quantifying the strength of neural tracking. Multiple comparisons can be naturally controlled by combining $p$-values using the harmonic mean~\citep{wilson2019harmonic}. However, testing based on $p$-values is not straightforward, since they are bounded and compressed near the tails. We therefore do not explore $p$-value-based comparison further here; instead, the following section introduces NNTS as a better alternative.

	\subsection{Comparing features via NNTS}
	\label{sec:use-case-d-prime}
	\citet{macintyre2026decoding} proposed using the Z-score of the raw correlations w.r.t. the null distribution as a metric to compare features. Here, we use the same idea, but based on the Fisher-transformed correlations. Given that the Z-score assumes normality of the underlying variables, it is the natural metric to combine with the Fisher-transformed correlations, which follow the normal distribution as shown in Section~\ref{sec:modeling}. To emphasize that this neural tracking metric contextualizes the correlations with their null distribution, we dub it the ``null-normalized tracking score'' (NNTS). Given $m$ Fisher-transformed neural tracking correlations $\{z_i = \artanh\!\left(r_i\right)\}_{i=1}^{m}$ with sample variance $\hat{\sigma}_{z}^2$ and null distribution variance $\hat{\sigma}_z^{(\text{null})^2}$, the window-level NNTSw and participant-level NNTSp are defined as\footnote{Note that we here implicitly assume that the real correlations too, with mean different from zero, can be modeled using the normal distribution after the Fisher transform. Given the theoretical explanations in Section~\ref{sec:modeling-fisher} and the verifications on null correlations in Section~\ref{sec:modeling-global-assessment}, we consider this reasonable. Furthermore, this modeling technique has already been successfully employed in~\citep{lopez2025unsupervised,geirnaert2025performance}.}:
	\begin{equation}
		\label{eq:nnts}
		\begin{cases}
			\text{NNTSw}_i = \frac{z_i}{\hat{\sigma}_{z}^{(\text{null})}}, & \text{(window-level)} \\
			\\
			\text{NNTSp} = \frac{\hat{\mu}_z}{\sqrt{\frac{1}{2}\left(\hat{\sigma}_z^{(\text{null})^2}+\hat{\sigma}_{z}^2\right)}}, & \text{(participant-level)}
		\end{cases}
	\end{equation}
	with $\hat{\mu}_z = \frac{1}{m} \sum_{i = 1}^m z_i, \hat{\sigma}_{z}^2 = \frac{1}{m-1} \sum_{i=1}^m (z_i-\hat{\mu}_z)^2$. Note that the denominator differs between the window- and participant-level definitions of NNTS, such that $\text{NNTSp} \neq \frac{1}{m}\sum_{i = 1}^m \text{NNTSw}_i$, unless $\hat{\sigma}_z^{(\text{null})^2} \approx \hat{\sigma}_{z}^2$ (which is often the case). The window-level NNTSw only references the null variance, keeping it a local, per-window metric. The participant-level NNTSp, on the other hand, uses the pooled standard deviation between the null and real distributions, and therefore differs from the averaged NNTSw. In fact, NNTSp is defined this way precisely because it becomes equal to d-prime (if $\hat{\sigma}_z^{(\text{null})^2} = \hat{\sigma}_{z}^2$, in theory), a signal detectability/sensitivity metric already used in measuring neural tracking (e.g., \citet{diLiberto2019low,crosse2021linear}). Similar to d-prime, this implies that NNTSp is a measure of distributional separability that also accounts for the variability in real correlations across windows. Both NNTSw and NNTSp, as well as the conversion to match-mismatch accuracy (See Section~\ref{sec:mm-acc}), are implemented in~\citep{geirnaert2026correlationToolbox}.

	\begin{figure} 
		\centering
		\begin{subfigure}{0.9\linewidth}
			\centering
			\includegraphics[width=1\linewidth]{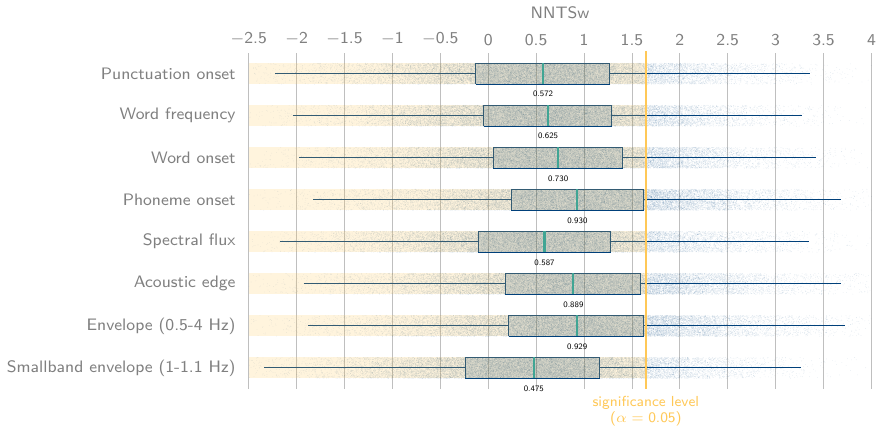}
			\caption{}
			\label{fig:useCase-nntsw}
		\end{subfigure}

		\begin{subfigure}{0.9\linewidth}
			\centering
			\includegraphics[width=1\linewidth]{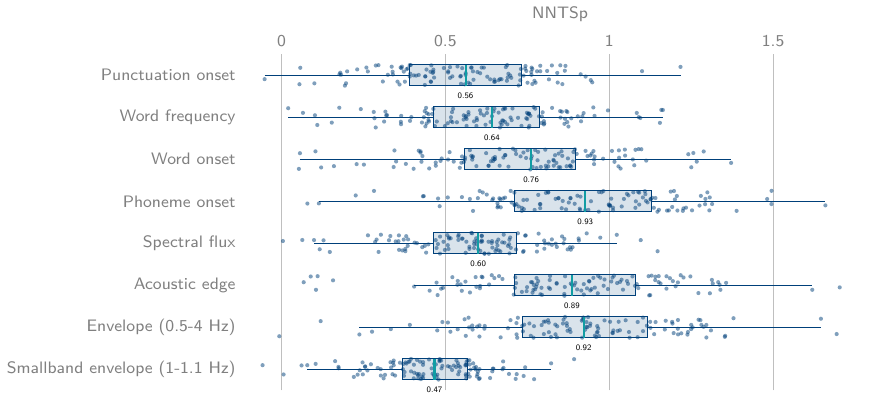}
			\caption{}
			\label{fig:useCase-nntsp}
		\end{subfigure}
		\caption{\textbf{(a)} The null-normalized tracking score (NNTS) at the window level (NNTSw) with $\SI{5}{\second}$ windows allows comparing features on the same scale and with a uniform significance level. \textbf{(b)} The NNTS at the participant level (NNTSp) summarizes neural tracking per participant and, again, allows comparing features on the same scale.}
		\label{fig:useCase_nnts}
	\end{figure}

	Figure~\ref{fig:useCase-nntsw} shows the NNTSw across features at the window level, and is therefore comparable with Figure~\ref{fig:useCase-raw-correlations-withSI}. Here, however, the NNTSw values are directly comparable across features, since normalizing with respect to the null distribution places them on the same scale. This also implies that the significance level, here for $\alpha = 0.05$, lies at the same value for every feature, i.e., $\text{NNTSw} = 1.645$, the $95\%$-percentile of the standard normal distribution $\mathcal{N}\!\left(0,1\right)$. It is now clear that the smallband envelope is the worst feature. Furthermore, the envelope, acoustic edge, and phoneme onset seem to perform on par, despite the acoustic edge showing considerably lower raw correlations in Figure~\ref{fig:useCase-raw-correlations-withSI}. The reason is clear from Figure~\ref{fig:useCase-raw-correlations-withSI}: the significance level is also lower for the acoustic edge. The NNTS has the advantage that it combines all of these considerations into a single number, facilitating easy comparison between features (and models or conditions, if necessary). A similar conclusion can be drawn from Figure~\ref{fig:useCase-nntsp}, showing the NNTS at the participant level.

	If desired, the NNTSp (shown in Figure~\ref{fig:useCase-nntsp}) can also be used for statistical testing. Here, just as an example, we run a linear mixed-effects model with feature as a fixed effect and a random intercept per participant, fitted by REML (lme4, R). A type-III ANOVA with Satterthwaite degrees of freedom was used as an omnibus test, showing a significant main effect of feature ($F = 354.87, p < 0.0001$). Given that we are not interested here in any particular comparison and use this merely as a showcasing example, we also run post-hoc pairwise comparisons across all features with Tukey multiple-comparisons correction (emmeans, R). The results in Table~\ref{tab:pairwise_nntsp} show that phoneme onset, the envelope, and acoustic edge form a top-performing tier, each significantly outperforming all remaining features (all $p < 0.0001$, $d > 1.3$). Within this tier, the pairwise comparisons are not fully transitive: the envelope does not significantly differ from either phoneme onset ($p = 0.9950$, $d = 0.10$) or acoustic edge ($p = 0.1475$, $d = 0.34$), yet phoneme onset and acoustic edge do differ significantly from each other ($p = 0.0168$, $d = 0.44$). Word onset forms its own, statistically distinct tier, falling below the top three (all $p < 0.0001$, $d > 1.3$) but above the remaining features (all $p < 0.0001$, $d > 1.0$). Among spectral flux, word frequency, and punctuation onset, effect sizes are generally small ($d < 0.4$ for two of the three pairwise comparisons), although word frequency and punctuation onset still differ significantly ($p < 0.0001$, $d = 0.71$). Finally, the smallband envelope is now clearly the worst-performing feature, significantly underperforming all others (all $p < 0.0001$, $d > 0.9$).

	\begin{table}[htbp]
		\centering
		\begin{tabular}{l l c c c c c}
		\toprule
		\multicolumn{2}{c}{Comparison} & Estimate & 95\% CI & $t(840)$ & $p$ (Tukey) & Cohen's $d$ \\
		\midrule
		Phoneme onset        & Envelope & 0.0101 & [$-$0.030, 0.051] &  0.76 & .9950     & 0.10 \\
		Phoneme onset        & Acoustic edge        & 0.0452 & [0.005, 0.086]    &  3.39 & .0168     & 0.44 \\
		Phoneme onset        & Word onset           & 0.1846 & [0.144, 0.225]    & 13.83 & $<.0001$ & 1.78 \\
		Phoneme onset        & Word frequency       & 0.2913 & [0.251, 0.332]    & 21.82 & $<.0001$ & 2.81 \\
		Phoneme onset        & Spectral flux        & 0.3308 & [0.290, 0.371]    & 24.79 & $<.0001$ & 3.19 \\
		Phoneme onset        & Punctuation onset    & 0.3645 & [0.324, 0.405]    & 27.31 & $<.0001$ & 3.51 \\
		Phoneme onset        & Smallband envelope   & 0.4636 & [0.423, 0.504]    & 34.74 & $<.0001$ & 4.47 \\
		Envelope & Acoustic edge        & 0.0351 & [$-$0.005, 0.076] &  2.63 & .1475     & 0.34 \\
		Envelope & Word onset           & 0.1744 & [0.134, 0.215]    & 13.07 & $<.0001$ & 1.68 \\
		Envelope & Word frequency       & 0.2811 & [0.241, 0.322]    & 21.07 & $<.0001$ & 2.71 \\
		Envelope & Spectral flux        & 0.3207 & [0.280, 0.361]    & 24.03 & $<.0001$ & 3.09 \\
		Envelope & Punctuation onset    & 0.3543 & [0.314, 0.395]    & 26.55 & $<.0001$ & 3.41 \\
		Envelope & Smallband envelope   & 0.4535 & [0.413, 0.494]    & 33.98 & $<.0001$ & 4.37 \\
		Acoustic edge        & Word onset           & 0.1393 & [0.099, 0.180]    & 10.44 & $<.0001$ & 1.34 \\
		Acoustic edge        & Word frequency       & 0.2461 & [0.205, 0.287]    & 18.44 & $<.0001$ & 2.37 \\
		Acoustic edge        & Spectral flux        & 0.2856 & [0.245, 0.326]    & 21.40 & $<.0001$ & 2.75 \\
		Acoustic edge        & Punctuation onset    & 0.3192 & [0.279, 0.360]    & 23.92 & $<.0001$ & 3.08 \\
		Acoustic edge        & Smallband envelope   & 0.4184 & [0.378, 0.459]    & 31.35 & $<.0001$ & 4.03 \\
		Word onset           & Word frequency       & 0.1067 & [0.066, 0.147]    &  8.00 & $<.0001$ & 1.03 \\
		Word onset           & Spectral flux        & 0.1462 & [0.106, 0.187]    & 10.96 & $<.0001$ & 1.41 \\
		Word onset           & Punctuation onset    & 0.1799 & [0.139, 0.220]    & 13.48 & $<.0001$ & 1.73 \\
		Word onset           & Smallband envelope   & 0.2791 & [0.238, 0.320]    & 20.91 & $<.0001$ & 2.69 \\
		Word frequency       & Spectral flux        & 0.0395 & [$-$0.001, 0.080] &  2.96 & .0621     & 0.38 \\
		Word frequency       & Punctuation onset    & 0.0732 & [0.033, 0.114]    &  5.48 & $<.0001$ & 0.71 \\
		Word frequency       & Smallband envelope   & 0.1723 & [0.132, 0.213]    & 12.91 & $<.0001$ & 1.66 \\
		Spectral flux        & Punctuation onset    & 0.0337 & [$-$0.007, 0.074] &  2.52 & .1877     & 0.32 \\
		Spectral flux        & Smallband envelope   & 0.1328 & [0.092, 0.173]    &  9.95 & $<.0001$ & 1.28 \\
		Punctuation onset    & Smallband envelope   & 0.0992 & [0.059, 0.140]    &  7.43 & $<.0001$ & 0.96 \\
		\bottomrule
		\end{tabular}
		\caption{Pairwise comparisons of NNTSp across features using Tukey multiple comparisons correction, ordered by descending feature mean.}
		\label{tab:pairwise_nntsp}
	\end{table}

	\subsubsection{Connection with match-mismatch accuracy}
	\label{sec:mm-acc}
	\citet{decheveigne2018decoding,deCheveigne2021auditory} put forward the match-mismatch task and its corresponding accuracy as a metric to compare models (or features) in neural tracking of speech. This match-mismatch task, which has been used in several other papers to compare features and models (e.g., \citet{diLiberto2019low,accou2021predicting,puffay2023relating,yao2023identifying,monesi2025auditory,diliberto2026robust}), consists of deciding whether a particular segment of the stimulus (feature) and neural response are temporally aligned or not. \citet{deCheveigne2021auditory} consider the match-mismatch accuracy a complementary metric to the correlation and the sensitivity index, akin to the NNTSp/d-prime metric defined in Equation~\eqref{eq:nnts}.

	Here, we close the circle by showing that the match-mismatch accuracy is simply one transformation away from the NNTSp metric. While neural tracking correlations have value, we have shown throughout this paper that careful interpretation with respect to the modeled null distribution is paramount. To facilitate comparison between features and models, we therefore proposed the NNTS metric. Because NNTS is built on the modeled null distribution via Algorithm~\ref{algo:siLevel-prediction} - which itself relies on the misalignment method - a direct connection with match-mismatch accuracy is to be expected, since match-mismatch accuracy is itself based on comparing aligned with misaligned segments of neural responses and stimulus features.

\begin{figure}
		\centering
		\begin{subfigure}{0.46\linewidth} 
			\centering
			\includegraphics[width=1\linewidth]{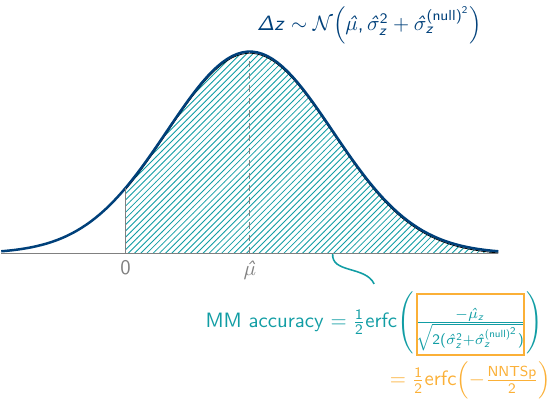}
			\caption{}
			\label{fig:MM-dist}
		\end{subfigure}%
		\begin{subfigure}{0.54\linewidth} 
			\centering
			\includegraphics[width=1\linewidth]{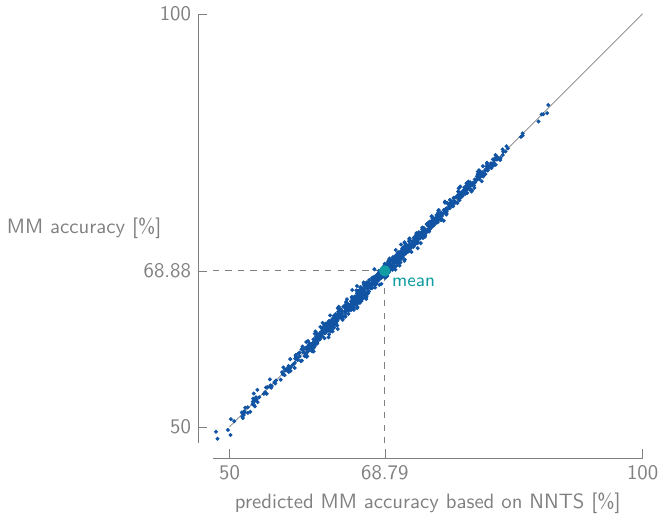}
			\caption{}
			\label{fig:sim-MMacc}
		\end{subfigure}
		\caption{\textbf{(a)} The match-mismatch decision variable $\Delta z$ follows a normal distribution with parameters derived from the modeled real and null distributions, enabling a direct transformation from NNTSp to the match-mismatch accuracy metric. \textbf{(b)} The NNTS-predicted match-mismatch (MM) accuracy at $\SI{5}{\second}$ windows across all features and participants closely corresponds with the real, measured MM accuracy.}
		\label{fig:MM-NNTS}
	\end{figure}

	Mathematically, this can be shown using a similar methodology as in modeling and predicting accuracies in selective auditory attention decoding~\citep{lopez2025unsupervised,geirnaert2025performance}. Considering the Fisher-transformed actual correlations $z = \artanh\!\left(r\right) \sim \mathcal{N}\!\left(\hat{\mu}_z,\hat{\sigma}_{z}^2\right)$ and misaligned null correlations $z^{(\text{null})} = \artanh\!\left(r^{(\text{null})}\right) \sim \mathcal{N}\!\left(0,\hat{\sigma}_{z}^{(\text{null})^2}\right)$, we have therefore already modeled the matched and mismatched correlations. As $\artanh$ is a monotonically increasing function, the match-mismatch decision rule becomes $\arg\!\max\!\left(r,r^{(\text{null})}\right) = \arg\!\max\!\left(z,z^{(\text{null})}\right)$. Assuming independence between $z$ and $z^{(\text{null})}$\footnote{While this is not necessarily true, as the underlying neural responses and stimuli are shared, the results in Figure~\ref{fig:sim-MMacc} show that, in practice, it is defensible.}, the match-mismatch decision variable can therefore equivalently be defined as:
	\[
	\Delta z = z-z^{(\text{null})} \sim \mathcal{N}\!\left(\hat{\mu}_z,\hat{\sigma}_{z}^2+\hat{\sigma}_{z}^{(\text{null})^2}\right) \lessgtr 0.
	\]
	As visualized in Figure~\ref{fig:MM-dist}, the match-mismatch accuracy can be easily derived from the normal distribution of $\Delta z$, with known parameters derived from the modeled real (matched) and null (mismatched) distributions:
	\[
	\begin{split}
		\text{match-mismatch accuracy} & = P\left(\Delta z > 0\right) \\
			& = \frac{1}{2}\text{erfc}\!\left(\frac{-\hat{\mu}_z}{\sqrt{2\left(\hat{\sigma}_{z}^2+\hat{\sigma}_{z}^{(\text{null})^2}\right)}}\right)\\
			& = \frac{1}{2}\text{erfc}\!\left(-\frac{\text{NNTSp}}{2}\right).
	\end{split}
	\]

	This shows that there is a direct mathematical relationship between the NNTSp (d-prime) metric and match-mismatch accuracy. This correspondence is empirically verified in Figure~\ref{fig:sim-MMacc}, comparing the predicted (NNTSp-based) and real match-mismatch accuracy at $\SI{5}{\second}$ windows. To empirically compute the real match-mismatch accuracy, for every feature and participant, the correlation between the neural response and the actual stimulus window is compared with the correlation against a mismatched stimulus window; the proportion of windows for which the matched correlation exceeds the mismatched one is defined as the match-mismatch accuracy. Across all participants and features, there is a close correspondence between the predicted and actual match-mismatch accuracy, with an overall $0.37$ percentage point absolute error, confirming that NNTS and match-mismatch accuracy carry the same information about the underlying distributional separability. Given the well-founded popularity of match-mismatch accuracy, as it implicitly accounts for the nature of the model, stimulus feature(s), SNR, \dots{} when evaluating performance, this mathematical relationship supports the validity of NNTS as a better-suited performance metric for quantifying neural tracking of natural stimuli than raw Pearson correlation coefficients, especially when comparing models and/or stimulus features. Different ways of choosing the mismatched segments, as proposed in~\citet{puffay2023relating}, then simply correspond to different criteria for performing the misalignment used to generate null correlations. Lastly, the extrapolation of the null variance (Equation~\eqref{eq:extrapolated-variance}) and of the real correlation distribution's variance, following the procedure in~\citet{geirnaert2025performance}, can be readily used to predict the match-mismatch accuracy across window lengths, by correspondingly extrapolating the variance in the computation of the NNTSp (Equation~\eqref{eq:nnts}).

	Although the two metrics are equivalent, NNTS is preferable in practice. The match-mismatch accuracy is a bounded proportion and therefore saturates: once neural tracking is strong or at long window lengths, accuracies compress towards $100\%$ and become insensitive to further differences between features or models, whereas NNTS is unbounded and continues to resolve them. The empirical accuracy moreover counts binary comparisons and thereby discards the magnitude of the underlying correlations, while NNTSp exploits the full continuous information. Finally, being unbounded and approximately normally distributed, it is also better suited to the parametric statistical testing illustrated above than a bounded proportion.

\begin{figure} 
		\centering
		\includegraphics[width=0.625\linewidth]{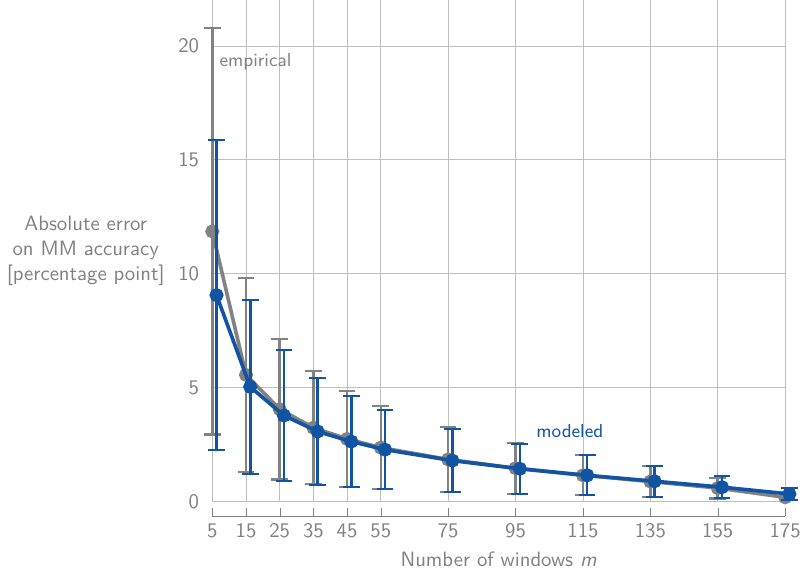}
		\caption{The mean $\pm$ SD (across participants, features, and $100$ resamplings) absolute error on the match-mismatch accuracy, obtained either through the modeled NNTSp or empirically, as a function of the number of windows $m$. The modeled approach is more accurate for small $m$, with both converging for larger $m$.}
		\label{fig:MMacc-nbWindows}
	\end{figure}

	This more efficient use of the available information also makes the \emph{modeled} match-mismatch accuracy based on NNTSp more reliable when few windows are available. Figure~\ref{fig:MMacc-nbWindows} compares the absolute error across features, participants, and $100$ resamplings on the match-mismatch accuracy obtained through NNTSp with that of the empirical estimate, as a function of the number of windows $m$. Each of the $m$ windows is compared with $m-1$ mismatches, while correspondingly $m(m-1)$ null correlations are used to compute the NNTSp. For small $m$, the modeled estimate is clearly more accurate ($9.1$ versus $11.9$ percentage points at $m = 5$), as the empirical proportion is estimated from a limited number of binary comparisons and is correspondingly more variable and quantized. This advantage gradually shrinks as $m$ grows, with both approaches performing equivalently from roughly $m = 100$ windows onward. The residual error of the modeled approach at the largest $m$ reflects its (small) model bias, consistent with the $0.37$ percentage point agreement reported above. Deriving the match-mismatch accuracy from NNTSp, if desired, is therefore particularly useful for short recordings, where the empirical accuracy becomes unreliable.

	\begin{highlightbox}{Key take-away \#5}{}
		Based on the modeled (null) distributions, raw neural tracking correlations can be more reliably interpreted. Features and models can be directly compared using the null-normalized tracking score (NNTS), which has a direct connection with the match-mismatch accuracy, yet is continuous and unbounded, and yields more reliable estimates when only limited data is available.
	\end{highlightbox}

	\section{Conclusion}
	\label{sec:conclusion}
	In this paper, we have provided a comprehensive framework to interpret and compare correlations between features and models in neural tracking of natural stimuli such as speech, music, and video. This framework consists of efficient and accurate modeling of null distributions and significance levels, and a new metric to more accurately quantify neural tracking, i.e., the null-normalized tracking score (NNTS). Interpreting neural tracking correlations in light of the null distribution and correct significance level is crucial, as the correlation can differ substantially in nature across, e.g., features and models. To facilitate adoption of the framework, all proposed methods are implemented in an accompanying toolbox~\citep{geirnaert2026correlationToolbox}.

	First, we showed that different permutation techniques for generating null correlations are not assumption-free: each implicitly assumes different properties of the underlying signals. We adopted misalignment as the default method for generating null correlations, arguing that it operationalizes the null hypothesis appropriate for content-specific tracking. Second, to address the need for generating many null correlations, we showed that neural tracking correlations can be well modeled using the normal distribution after the Fisher transform. This semi-parametric model enables accurate estimation of the significance level. Moreover, because the null distribution is well modeled even from few null correlations, it enables \emph{efficient} significance-level estimation. The semi-parametric model yields better significance-level estimates than the empirical null distribution for fewer than $\num{10 000}$ permutations, with excellent approximations for as low as $\num{1000}$ permutations. Third, the significance level can be reliably extrapolated from one window length to another using the model, without needing to recompute null correlations at different window lengths, greatly reducing computational cost. Combined, we established that, using $\SI{5}{\second}$ windows, $3-\SI{5}{\minute}$ of data suffices to reliably estimate the null distribution and significance level.

	Beyond this null-distribution and significance-level methodology, we also showed how features can be interpreted using the significance level at the window and participant level. To enable direct comparison of features and models on the same scale, we proposed the null-normalized tracking score (NNTS) and demonstrated its use in statistical testing of neural tracking. Moreover, we mathematically and empirically showed that NNTS has a direct relationship with the often-used match-mismatch accuracy and that deriving it from NNTS is more reliable when few windows are available, providing a coherent framework to interpret and compare features and models in neural tracking, beyond simply - and incorrectly - using raw correlations as a performance metric.

	The validation and use case in this paper were performed on a single dataset, using backward modeling of speech features in normal-hearing adults. The proposed statistical framework itself is agnostic to stimulus modality and recording technique, and we see no specific reason it would not extend to other naturalistic stimuli (e.g., music, video) or modalities (e.g., MEG, ECoG). However, this generalization remains to be empirically demonstrated.

	\section*{Data and Code Availability}
	The raw EEG data are part of the publicly available SparrKULee dataset~\citep{accou2024sparrkulee}; data from the additional participants included here could not be made publicly available. The neural tracking correlations and null correlations derived from these recordings, which underlie all analyses from Section~\ref{sec:modeling} onward, are available at~\citet{geirnaert2026data}. All MATLAB and R code required to reproduce the results, figures, and tables from Section~\ref{sec:modeling} onward is available at~\citet{geirnaert2026experimentCode}. A toolbox implementing the proposed methodology, including the misalignment procedure, the semi-parametric null distribution model, significance level estimation and extrapolation, and the NNTS metric, is available at \url{https://github.com/Gonaz/NTC} \citep{geirnaert2026correlationToolbox}.

	\section*{Author Contributions}
	\textbf{Simon Geirnaert:} Conceptualization, Methodology, Software, Validation, Formal analysis, Investigation, Writing - Original Draft, Visualization, Project administration, Funding acquisition.
	\textbf{Alexander Bertrand:} Methodology, Writing - Review \& Editing, Supervision, Funding acquisition.
	\textbf{Tom Francart:} Writing - Review \& Editing, Supervision, Funding acquisition.
	\textbf{Jonas Vanthornhout:} Conceptualization, Methodology, Software, Formal analysis, Investigation, Writing - Original Draft, Data Curation, Funding acquisition.

	\section*{Funding}
	Financial support was provided by the Research Foundation Flanders (FWO) (junior postdoctoral fellowship fundamental research 1242524N for S. Geirnaert; project No G081722N and G026026N), Internal Funds KU Leuven (projects IDN/23/006, C14/25/108, C3/25/017), the European Research Council (ERC) under the European Union's research and innovation programe (grant agreement No 802895 and grant agreement 101138304), and the Flemish Government (AI Research Program). Views and opinions expressed are however those of the author(s) only and do not necessarily reflect those of the European Union or any of the granting authorities. Neither the European Union nor the granting authorities can be held responsible for them.

	\section*{Declaration of Competing Interests}
	The authors declare no competing interests.

	\section*{Acknowledgments}
	Claude Sonnet (version 4.6; Anthropic) was used as a writing support for language editing and grammar checking, and as a coding assistant for software development and implementation support. All text and code was originally written by the authors, and all algorithms, analyses, and interpretations were designed, verified, and validated by the authors, who take full responsibility for the results. All authors have reviewed and approved the final content of the manuscript and take full responsibility for the paper's scientific integrity.

	\bibliographystyle{elsarticle-harv}
	\bibliography{biblio}

	\appendix
	\section{Comparison of the Fisher transform with other modeling techniques}
	\label{app:comparison}
	We consider the following three alternatives to the normal distribution after the Fisher transform for modeling the null distribution of Pearson correlation coefficients between neural responses and stimulus features:
	\begin{description}
		\item[Normal distribution] The null correlations $r^{(\text{null})}$ can be modeled directly with a zero-mean normal distribution $r^{(\text{null})} \sim \mathcal{N}\!\left(0,\sigma_r^{(\text{null})^2}\right)$, with the zero mean following automatically from the null hypothesis and the variance estimated directly from the generated null correlations. This is loosely motivated by the central limit theorem, since the Pearson correlation coefficient involves a summation over many samples. Strictly, normality does not hold here: the time samples are not independent, and the correlation is bounded to $\left[-1,1\right]$. Nonetheless, it can be a convenient choice in practice, especially for long windows, where many samples are summed and the variance of the correlation is small.
		\item[Truncated normal distribution] To make the normal model theoretically more sound, the truncated normal distribution bounds it to $\left[-1,1\right]$. The mean remains zero, while the standard deviation is obtained by the method of moments, i.e., by equating the sample variance with the variance of the distribution~\citep{horrace2015moments}.
		\item[Student's t-distribution] Inspired by significance testing for Pearson correlation coefficients, the null distribution can also be modeled with a Student's t-distribution. As shown in \citet{kendall1979advanced}, under the transformation
		\[
		t = r\sqrt{\frac{\text{DOF}}{1-r^2}} \sim \text{t-distribution}\!\left(\text{DOF}\right),
		\]
		$t$ follows a Student's t-distribution when the underlying variables are normally distributed and uncorrelated, with DOF the degrees-of-freedom parameter. Here, we treat DOF as a parameter to be estimated, again by the method of moments after transformation.
	\end{description}
	
	We compare these methods against the normal distribution after the Fisher transform using the same significance-level estimation metrics as in Section~\ref{sec:modeling-si-levels}: the relative correlation error (Table~\ref{tab:model-comparison-correlationError}) and absolute percentile error (Table~\ref{tab:model-comparison-percentileError}). These per-feature errors, averaged across participants and window lengths, show that the normal distribution after the Fisher transform consistently outperforms all alternatives, most clearly for edge cases such as the smallband envelope and phoneme and punctuation onsets, and for shorter windows, where the distributions deviate most from normality.

	\begin{table}[]
		\centering
		\begin{subtable}{0.65\linewidth}
			\includegraphics[width=1\linewidth]{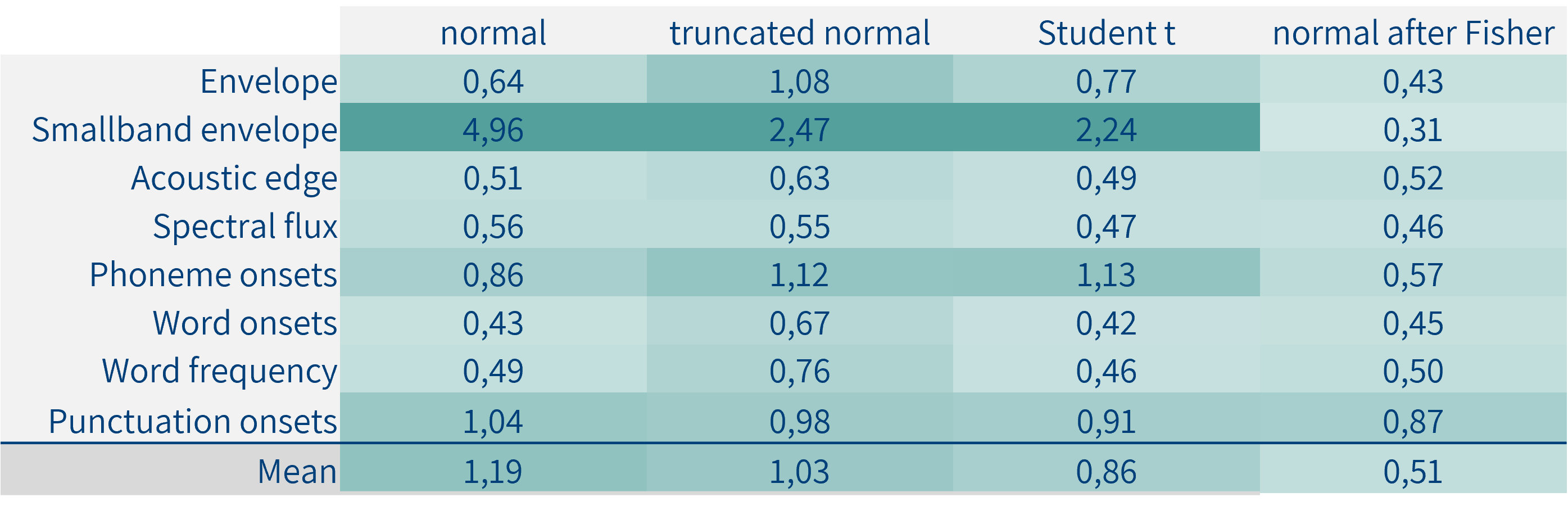}
			\caption{}
			\label{tab:model-comparison-correlationError}
		\end{subtable}

		\begin{subtable}{0.65\linewidth}
			\includegraphics[width=1\linewidth]{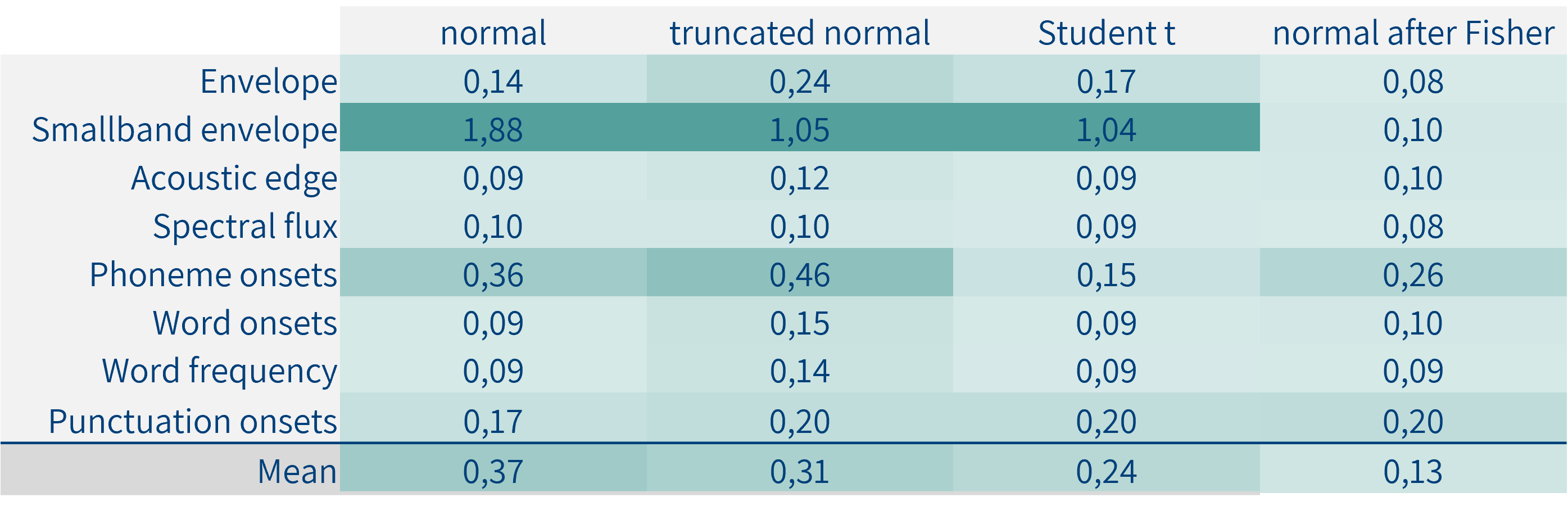}
			\caption{}
			\label{tab:model-comparison-percentileError}
		\end{subtable}
		\caption{The average (across participants and window lengths) \textbf{(a)} relative correlation error on the estimated $95\%$-significance level and \textbf{(b)} absolute percentile error on the nominal $\alpha = 0.05$ level. Both are lowest overall for the normal distribution after the Fisher transform, with the largest differences at shorter window lengths for the smallband envelope, phoneme onset, and punctuation onset.}
		\label{tab:model-comparison}
	\end{table}

	\section{Modeling the null distribution of across-window-averaged correlations}
	\label{app:model-avg-correlations}
	As opposed to single Pearson correlation coefficients, there is no rigorous characterization for averaged (null) correlation coefficients $\bar{r}_m^{(\text{null})} = \frac{1}{m} \sum_{i=1}^mr_i^{(\text{null})}$ across $m$ windows. The normal distribution after Fisher transform for single correlations does not offer a way out, as the hyperbolic tangent function does not commute with the sum. Therefore, the most relevant mechanism to characterize the distribution of $\bar{r}_m^{(\text{null})}$ seems to be central limit theorem, which states that its distribution should converge to a normal distribution. Given estimated variance of the per-window raw correlations (with mean assumed $0$):

	Unlike single Pearson correlation coefficients, averaged null correlations $\bar{r}_m^{(\text{null})} = \frac{1}{m} \sum_{i=1}^m r_i^{(\text{null})}$ across $m$ windows have no rigorous distributional characterization. The normal distribution after the Fisher transform, used for single correlations, offers no way out here, since the hyperbolic tangent does not commute with averaging: the Fisher transform of the mean correlation is not the mean of the Fisher-transformed correlations. The most natural remaining route is the central limit theorem, which - assuming the per-window null correlations are approximately independent - implies that $\bar{r}_m^{(\text{null})}$ converges to a normal distribution as $m$ grows. Given the estimated per-window null variance (with mean assumed zero),
	\[
		\hat{\sigma}_r^{(\text{null})^2} = \frac{1}{n} \sum_{i=1}^{n} r_i^{(\text{null})^2},
	\]
	the averaged null correlation is then modeled as
	\[
		\bar{r}_m^{(\text{null})} \sim \mathcal{N}\!\left(0,\frac{\hat{\sigma}_r^{(\text{null})^2}}{m}\right).
	\]
	This procedure is summarized in Algorithm~\ref{algo:avg-estimation}.

	We compare the estimated $95\%$-significance levels for the average neural tracking correlation on $\SI{5}{\second}$ windows, obtained via Algorithm~\ref{algo:avg-estimation}, against the empirical estimate, for $m$ ranging from $2$ to $175$ (the maximum number of $\SI{5}{\second}$ windows available for averaging). The `ground-truth' null distribution of average correlations is constructed by computing, $\num{10000}$ times, the average of $m$ correlations subsampled from the original $\num{100000}$ null correlations. For each $m$, participant, and feature, both the empirical and the modeled significance-level estimate are based on $\num{1000}$ randomly selected null correlations (based on the findings in Section~\ref{sec:estimating-si-levels}). For the empirical estimate specifically, the null distribution is built from $\num{10000}$ averages, each subsampled from these $\num{1000}$ selected correlations.

\begin{figure} 
		\centering
		\includegraphics[width=0.65\linewidth]{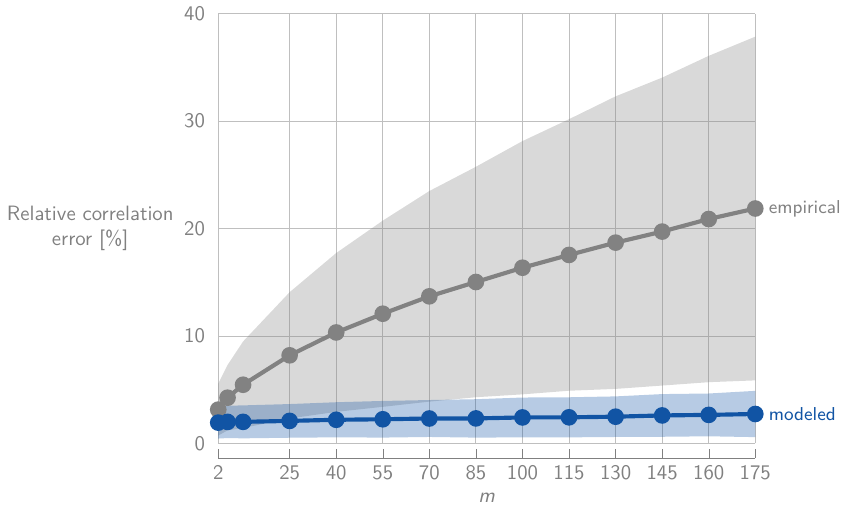}
		\caption{The average (across participants and features; shaded area: $\pm 1$ SD) relative correlation error on the estimated $95\%$-significance level of the average correlation as a function of the number of averaged windows $m$.}
		\label{fig:avgCorrelations-relError}
	\end{figure}

	Figure~\ref{fig:avgCorrelations-relError} shows the average relative correlation error on the $95\%$-significance level across participants and features. The modeled approach based on the normal distribution achieves low errors throughout, averaging $2.31\%$ across $m$. In contrast, the empirical estimate degrades substantially as $m$ grows, since fewer independent draws can be formed for larger $m$. While the model's relative errors are slightly higher than for individual correlations (cf.\ Figure~\ref{fig:result2_correlationError} at $\num{1000}$ permutations, with $1.61\%$ error), they remain low overall.

\end{document}